\documentclass[preprint,12pt]{elsarticle}

\usepackage{graphicx}
\usepackage{amssymb}
\usepackage{amsthm}

\usepackage{color}
\usepackage{ulem}
\usepackage{comment}

\def\clst{{\rm{clst}}}

\def\max{{\rm{max}}}

\def\inf{{\rm{inf}}}
\def\mat{{\rm{mat}}}

\def\Vol{{\rm{vol}}}
\def\agg{{\rm{agg}}}
\def\total{{\mathrm{total}}}
\def\cB{{\mathcal{B}}}
\def\cL{{\mathcal{L}}}

\def\Rconf{{\mathfrak{R}}}
\def\talpha{{t}}
\newcommand{\by}[1]{{\textrm{#1},}}
\newcommand{\jour}[1]{{\textrm{#1}}}

\newcommand{\vol}[1]{{\textbf{#1}}}
\newcommand{\yr}[1]{{(\textrm{#1})}}
\newcommand{\pages}[1]{{\textrm{#1}}}
\newcommand{\Book}[1]{{\textit{#1},}}
\newcommand{\publ}[1]{{(\textrm{#1},}}
\newcommand{\publaddr}[1]{{\textrm{#1},}}
\newcommand{\byr}[1]{{\textrm{#1})}}

\journal{Applied Mathematical Modeling}

\begin{document}

\begin{frontmatter}



\title{NUMERICAL COMPUTATIONS OF CONDUCTIVITY OVER
AGGLOMERATED CONTINUUM PERCOLATION MODELS}


\author[Mats]{Shigeki Matsutani}
\author[Shim]{Yoshiyuki Shimosako}
\author[Wang]{Yunhong Wang}

\address[Mats]{Analysis technology center, Canon Inc.\\
 3-30-2, Shimomaruko, Ohta-ku, Tokyo 146-8501, Japan\\
    matsutani.shigeki@canon.co.jp}

\address[Shim]{Analysis technology center, Canon Inc.\\
 3-30-2, Shimomaruko, Ohta-ku, Tokyo 146-8501, Japan\\
    shimosako.yoshiyuki@canon.co.jp}

\address[Wang]{Analysis technology center, Canon Inc.\\
 3-30-2, Shimomaruko, Ohta-ku, Tokyo 146-8501, Japan
\footnote{Present address:
Microcreate System Co.,
2-1-15 4C, Chuou Yamato Kanagawa 242-0021 Japan.}}

\begin{abstract}

In order to clarify how
the percolation theory governs the conductivities
 in real materials which consist of small conductive particles,
{\it{e.g.}}, nanoparticles, with random configurations
%\textcolor{blue}{
in an insulator,
%} 
we numerically investigate the conductivities of continuum percolation
models consisting of overlapped particles using the finite difference
method as a sequel of our previous article
(Int. J. Mod. Phys. {\bf{21}} (2010), 709).
As the previous article showed the shape effect of each particle
by handling different aspect ratios of spheroids, in this article
we numerically 
%\textcolor{red}{\sout{reveal}}
%\textcolor{blue}{
show
%}
 influences of the agglomeration of the particles
on conductivities after we model the agglomerated configuration
by employing a simple numerical algorithm which simulate
an agglomerated configuration of particles by a natural parameter.
We conclude that the dominant agglomeration effect on the conductivities
can be interpreted as the size effect of an analyzed region.
%\textcolor{blue}{
We also discuss an effect of shape of the agglomerated 
clusters on its universal property.
%}
\end{abstract}


\end{frontmatter}

% \linenumbers

\section{Introduction}

As in the previous article \cite{MSW}, we investigate conductivities
of a percolation model using the finite
difference method (FDM) \cite{LV}. The purpose of
this article and the previous one is to clarify how
the percolation theory governs the conductivities
 in real materials which consist of small conductive particles,
{\it{e.g.}}, nanoparticles, with random configurations 
%\textcolor{blue}{
in an insulator.
%} 
In general, the small particles in a real material have
some typical sizes,  shapes like spheroids, and
cluster structures due to their agglomeration \cite{AS}-\cite{MB}.
The purpose of this article is to investigate an agglomeration effect on
the conductivity whereas the previous one \cite{MSW} is a study on
the effect of the shape of each particle.
We propose a simple numerical algorithm to simulate an agglomerated
configuration of particles though there are some physical models
which numerically simulate the agglomerated phenomena \cite{TTK,KH}.
In this article, we employ the simple algorithm.
Since our algorithm has a parameter which controls the agglomerated
configurations consistently without any difficulties,
we numerically show how the conductivity depends upon the agglomerations.
Then we 
%\textcolor{red}{\sout{reveal}}
%\textcolor{blue}{
find
%}
the fact that the main agglomeration effect on
the conductivity can be interpreted as the size effect of the system.
%\textcolor{blue}{
We also discuss an effect of shape of the agglomerated 
clusters on the universal property of the conductivity.
%}

The conductivity is one of the most concerned phenomena
in the percolation theory \cite{SA}{-}\cite{Ba1}
and it is important to investigate the conductivity curve
\begin{equation}
        \sigma_\total = c (p - p_c)^\talpha,
 \label{eq:1-2}
\end{equation}
where $c$ is a constant factor, $p$ is the volume fraction,
$p_c$ is the percolation threshold, and $\talpha$ is the critical
exponent of the conductivity.
The behaviors of the threshold $p_c$ and the exponent $\talpha$
represent the conductive properties of the systems in the
percolation phenomena, which have some universal properties.
It means that their dependence on some parameters of a model
is sufficiently weak due to the randomness.
In order to discuss the 
%\textcolor{red}{\sout{universality}}
%\textcolor{blue}{
universal properties,
%}, 
the percolation theory is, 
%\textcolor{red}{\sout{ordinarly,}}
%\textcolor{blue}{
basically,
%}
based upon a system with the
infinite size \cite{SA}${}^{-}$\cite{G}.

However when we apply the percolation theory to the real composite
materials of conductive nanoparticles 
%\textcolor{blue}{
in an insulator,
%} 
we encounter effects coming
from the shape of the nanoparticles and several characteristics
lengths.  Besides lattice percolation models which are given on
discrete lattices, the continuum percolation model (CPM) was introduced
as in Refs.~\cite[p.108-111]{SA} and \cite{MR}. In order to
handle the size and the shape of particles, CPM has been 
studied
%\textcolor{blue}{
\cite{MR,GSDT}.
%}
In the previous article \cite{MSW}, we 
%\textcolor{blue}{
numerically
%}
studied the shape effects
on the conductivities in CPMs among the different aspect ratios
of spheroids, where we allowed overlap of the particles.
%\textcolor{blue}{
As a numerical method, we employed the finite difference method (FDM),
domain decomposition method on the parallel computations and
the preconditioned conjugate gradient method (PCGM)\cite{LV},
though similar attempts to estimate transport properties or
electric properties in CPM using the finite element method (FEM)
appeared 
in Refs.~\cite{Y,MBr,LWLS}. For a quite complicated geometrical
objects, FDM over the regular lattice is basically robust
from the viewpoint of the numerical computations. 
Then the
%} 
previous article shows that the conductivity in CPM strongly
depends upon the shape of the composed particles.

In this article, we will consider another shape effect or the agglomerated
effect of these stuffed spherical particles on the conductivity.
With the development of the technology, 
the smaller the size of the (nano-)particles becomes,
the smaller the size of the related devices becomes.
When we investigate the conductivities in the real composite materials
of conductive nanoparticles from the viewpoint of the percolation theory,
we handle three characteristic lengths, i.e., 1) the size of
the particle as a minimal size, 2) the size of the system, e.g.,
the thickness of the film, as the maximal size, and 3) the size of
the percolation cluster which is given by a multiplication of
the size of the particles and is also related to the size of the system.
When the size of devices is sufficiently small, it is important to
consider the size effect of the system though the conductivities
in the system with infinite size are basically concerned in the
percolation theory.
In other words, in order to apply the percolation theory to a real
material system, it is crucial to consider the relations among these scales.

The smaller the size of particles is,
the more agglomerated the particles becomes, due to their interface energy.
Agglomeration forms agglomerated clusters and the cluster brings the
fourth characteristic scale to the system.
Thus the evaluation of the agglomeration effect on the conductivity is
very important.
By employing the simple algorithm to simulate the agglomeration
of particles, we numerically give a series of agglomerated configurations
of CPM, which we call {\it{agglomerated continuum percolation models}}
(ACPMs), and investigate the conductive properties of ACPMs by means of FDM.
In this article, we also handle the overlapping particles \cite{MSW}.
We show the dependence of the conductivity curves on the agglomeration
as in Figure \ref{fig:thandexvsag}
%\textcolor{blue}{
since our algorithm provides natural properties
from view point of a conditional probabilistic problem as we show in
Subsection 2.1.
%}

Since the finite size effects in the conductivity on a percolation
model were studied in Chapters 4 and 5 of Ref.~\cite{SA}, based upon these
studies, we numerically consider the relation between the geometrical
effects and the properties of the conductivities in ACPMs.
As a result, we show that one of the dominant agglomerated effects on
the conductivities with ACPM can be regarded as the size effects
of the system. 
%\textcolor{red}{\sout{and the}}
%\textcolor{blue}{
We also show that it is expected that
%}
the shape of the agglomerated clusters 
%\textcolor{blue}{
might
%}
affect the conductivity.
The shape effect implies that the agglomeration would have
 an effect on the 
%\textcolor{red}{\sout{universality class}}
%\textcolor{blue}{
universal properties
%}
 of the conductivity curve
 (\ref{eq:1-2}), and 
%\textcolor{blue}{
thus we numerically
%\sout{show the possibility in Figure 
%\ref{fig:smallparticle}, though it not an evidence of the effect.
%}
%}
%\textcolor{blue}{
discuss the relation between the shape effect and the agglomeration
effect in Subsection 4.2.3.
%}

Contents in this article are as follows.
Section 2 shows our computational method.
Subsection 2.1 describes our algorithm to construct the agglomerated clusters
in CPM and the geometrical setting in ACPMs.
Section 2.2 provides the computational method of the conductivity
over ACPMs using FDM, which
is basically the same as that in the previous article \cite{MSW}.
Section 3 shows our computational results of the conductivity
in ACPMs with the agglomerated clusters.
In Section 4, we discuss our results from geometrical viewpoints.
In Section 5, we summarize the results and the discussions.

\section{Geometrical setting of ACPM}

In this section, we show our geometrical setting of ACPM. We model
the agglomerated clusters out of a simple algorithm which is governed by
a parameter $\gamma_\agg\in[0,1]$.
We also briefly show the computational method of the conductivities
over ACPMs using FDM in Subsection 2.2, whose details are described in  
the previous article \cite{MSW}.
%We first give the simple algorithm of ACPM as follows.

\subsection{Agglomeration algorithm}

We set particles parametrized by their positions $(x,y,z)$ into
a box-region $\cB:=[0,x_0]\times [0,y_0]\times [0,z_0]$ at random and
get a configuration $\Rconf_n$ as one of CPMs. In this article,
we set $x_0 = y_0 = z_0 = L$.  The particle corresponds to a stuffed
sphere or ball with the same radius $\rho$,
$B_{x_i, y_i, z_i}:=
\{ (x, y, z) \in \cB \ | \ |(x,y,z)-(x_i, y_i, z_i)| \le \rho\}$.
The configuration $\Rconf_n$ is given by 
$\Rconf_n:=\bigcup_{i=1}^n B_{x_i, y_i, z_i}$.

By fixing the radius $\rho$ of the particle, $\rho = 1$, and a number
$\gamma_\agg\in [0,1]$ which is called agglomeration parameter, we
introduce an algorithm to construct the configuration $\Rconf_n$ in ACPM.
%%SHIM_deleted
%\bigskip
%%SHIM_deleted

\begin{figure}[htbp]
\includegraphics[scale=0.6]{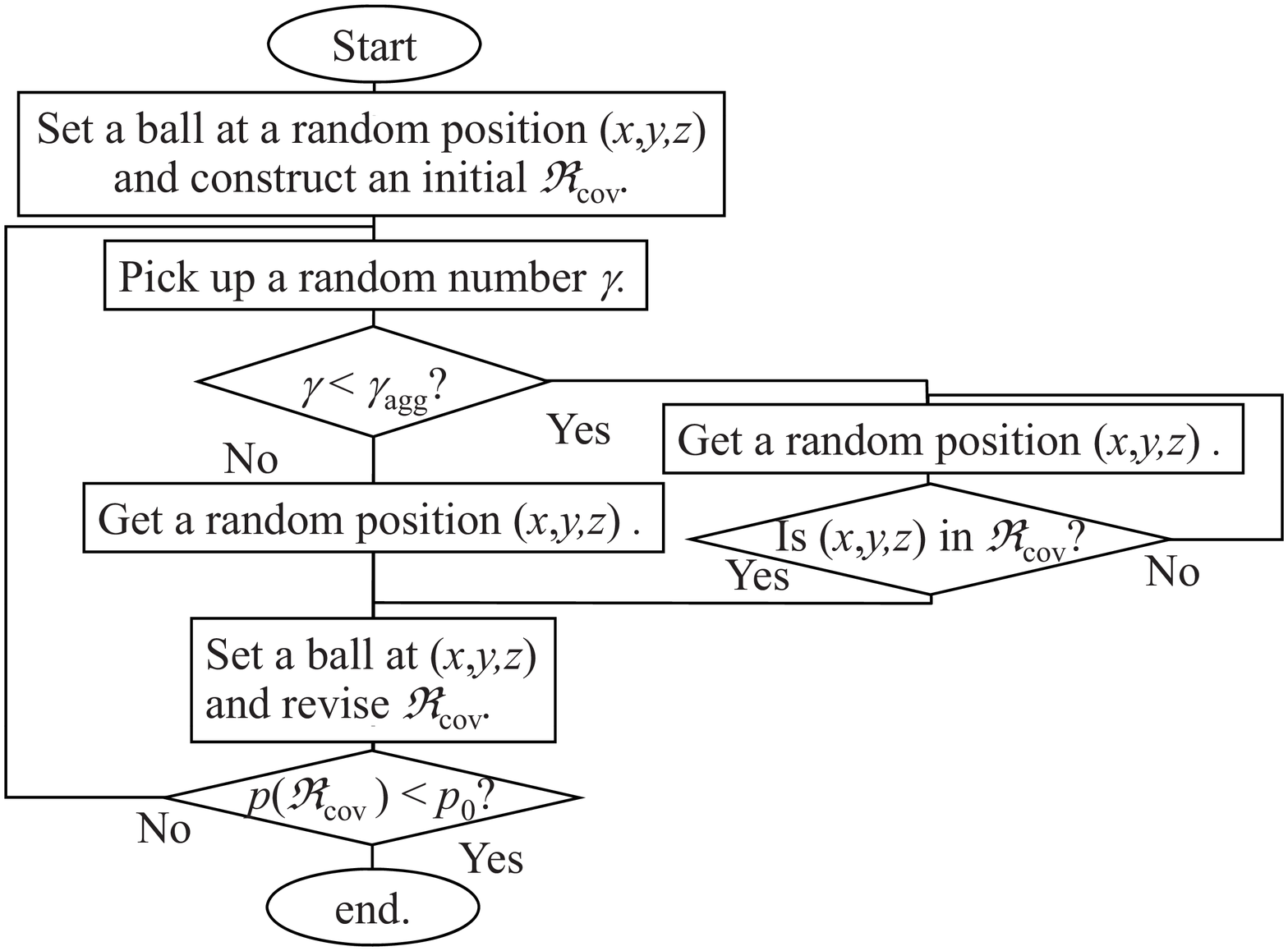}
\caption{\label{fig:flowchart} The flowchart of the agglomeration
configuration algorithm.}
\end{figure}
We illustrate our algorithm by a flowchart in Figure \ref{fig:flowchart}.
As an initial state, the configuration $\Rconf_0$ has no particle.
As the first step, for a uniform random position $(x,y,z) \in \cB$
we set a particle $B_{x,y,z}$ whose center is $(x,y,z)$ and
the radius is $\rho$, i.e., $\Rconf_1:=B_{x,y,z}$.

Let us consider $n$-step. We take a position $(x,y,z)$ at uniform random
in $\cB$, and another random parameter $\gamma$ at uniform random
in $[0,1]$.  If the random parameter $\gamma$ is greater than $\gamma_\agg$,
we employ the position $(x,y,z)$ as the center of a particle and
$\Rconf_{n+1}:=\Rconf_n \bigcup B_{x,y,z}$.
It is noted that we allow the particles to overlap each other.

When the random parameter $\gamma$ is given as $\gamma \le \gamma_\agg$,
we first check whether the ball whose center is the position $(x,y,z)$
is connected with the previous configuration $\Rconf_n$  or not. 
If it is connected with the configuration $\Rconf_n$, or 
$\Rconf_n\bigcap B_{x,y,z}\neq \emptyset$, we employ the
position and add the particle into the configuration $\Rconf_n$, or
$\Rconf_{n+1}:=\Rconf_n \bigcup B_{x,y,z}$.

If not, i.e., $\Rconf_n\bigcap B_{x,y,z}= \emptyset$, 
we abandon the position and go on to take another
uniformly random position $(x,y,z)$ in $\cB$ until we find 
the position which supplies a connected particle $B_{x,y,z}$ with $\Rconf_n$.

In other words, if we take $\gamma$ which is smaller than $\gamma_\agg$,
the added particle must be connected with the previous configuration 
$\Rconf_n$. Thus, $\gamma_\agg$ stands for the agglomeration of the 
particle system. 

%We  put the ball one by one until the total 
%olume fraction of $\sigma_\mat$ in $\cB$ became a certain value by monitoring
%he number of the cells which had $\sigma_\mat$.
By monitoring the total volume fraction which is
a function of $\Rconf_n$ and is denoted by $\Vol(\Rconf_n)$, 
we continue to put the particles as long as
$\Vol(\Rconf_n) \le p$ for a given volume fraction $p$.
We find the step $n(p)$ such that
$\Vol(\Rconf_{n(p)-1}) \le p$ and
$\Vol(\Rconf_{n(p)}) > p$.
Since the difference between $\Vol(\Rconf_{n(p)-1})$ 
and $\Vol(\Rconf_{n(p)})$ is at most $0.9\times 10^{-4}$  
for the employed parameters,
we regard $\Vol(\Rconf_{n(p)})$ as the volume fraction $p$ itself
hereafter under this accuracy.

%%SHIM_deleted
%\bigskip
%%SHIM_deleted
\begin{figure}[htbp]
\includegraphics[scale=0.5]{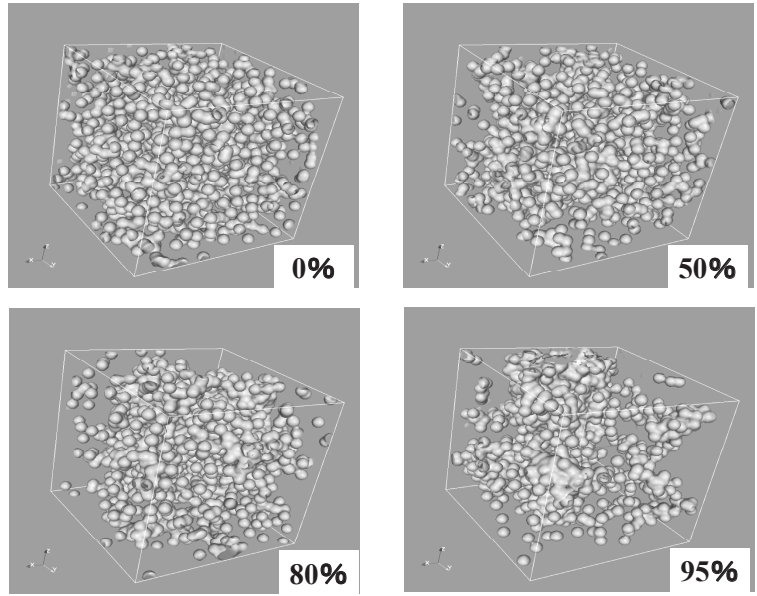}
\caption{\label{fig:3Dview} The agglomeration parameters:
Random particle systems with the agglomeration parameter $\gamma_\agg$
$=0.0, 0.5, 0.8$ and $0.95$ for (a), (b), (c) and (d) respectively.}
\end{figure}
As illustrated in Figure \ref{fig:3Dview}, the uniform random configuration
is realized as the case $\gamma_\agg = 0$, and the agglomeration
configurations are given with our agglomeration parameter $\gamma_\agg > 0$.
We regard  ACPM of the case $\gamma_\agg=0$ as 
the ordinary CPM for balls with the same radius \cite{SA,MR}
and thus we simply call it CPM  hereafter.

%%SHIM_deleted
%\bigskip
%%SHIM_deleted
%\textcolor{blue}{
In our algorithm,  
$B_{x,y,z}$ in $\Rconf_n$ is 
determined only by the previous configuration $\Rconf_{n-1}$ as 
a conditional probabilistic problem like $\Pr(B_{x,y,z}|\Rconf_{n-1})$.
Since the configuration
$\Rconf_{n-1}$ memorizes $n-1$ past outcomes  
$(B_{x_i,y_i,z_i})_{i=1,2,\ldots,n-1}$,
it cannot be regarded as a Markov chain for $\gamma_\agg \neq 0$ 
\cite{Fel,Kol}.
However we have a natural hierarchical properties
$\Rconf_{n} \subset \Rconf_{n+1}$
as a sequential events governed by conditional probabilities
for $\gamma_\agg\neq 0$ case
and by the independent probability for the 
vanishing $\gamma_\agg$ case.
%}
Since we use the pseudo-randomness to simulate the random
configuration $\Rconf_{n(p)}$ for given $p$ and $\gamma_\agg$,
the configuration $\Rconf_{n(p)}$ depends upon the
 seed $i_s$ of the pseudo-randomness which we choose.
We let it be denoted by $\Rconf_{\gamma_\agg, p, i_s}$.
%\textcolor{blue}{
In other words, for a certain 
seed $i_s$ of the pseudo-randomness,
 we handle the
set of the configurations 
$\{\Rconf_{\gamma_\agg, p, i_s} \ | \ p \in [0,1]\}$ 
as the sequence of the outcomes of the conditional probabilistic problems.
Due to the hierarchical properties,
$\Rconf_{\gamma_\agg, p, i_s}$ is the events under 
the prior conditions $\Rconf_{\gamma_\agg, p', i_s}$ for $p' < p$,
%}
%\textcolor{red}{\sout{
%For the same seed $i_s$ of the pseudo-randomness, a}
%}
%\textcolor{blue}{
and the
%}
 configuration $\Rconf_{\gamma_\agg, p,i_s}$ of a volume fraction $p$ 
naturally contains a configuration $\Rconf_{\gamma_\agg, p', i_s}$ 
for $p'<p$, i.e.,
 $\Rconf_{\gamma_\agg, p',i_s} \varsubsetneqq \Rconf_{\gamma_\agg, p,i_s}$.
Hence the elements in the set of the configurations 
$\{\Rconf_{\gamma_\agg, p,i_s}\ | \ p \in [0,1]\}$ 
keeping the same seed $i_s$ are  relevant
%\textcolor{blue}{
and have the (total ordered) hierarchical structure,
\begin{equation}
 \Rconf_{\gamma_\agg, 0,i_s} 
\varsubsetneqq \Rconf_{\gamma_\agg, p_1,i_s}
\varsubsetneqq \Rconf_{\gamma_\agg, p_2,i_s}
\varsubsetneqq \cdots 
\varsubsetneqq \Rconf_{\gamma_\agg, 1,i_s},
\label{eq:FiltR}
\end{equation}
for $0 < p_1 < p_2< \cdots < 1$.
%}
%\textcolor{red}{\sout{
%As shown later, the conductivity curve (\ref{eq:1-2})
%over  $\{\Rconf_{\gamma_\agg, p,i_s}\ | \ p \in [0,1]\}$ 
%for fixing $i_s$ is}}
%\textcolor{blue}{well-defined as a function over a measure space.}
%\textcolor{red}{\sout{
%obtained
%as a smooth curve even for the non-vanishing $\gamma_\agg$}}.
%
%
%
%\textcolor{blue}{
%As mentioned above, our algorithm generates a family of
%configurations $\{\Rconf_{\gamma_\agg, p, i_s} \ | \ p \in [0,1]\}$ 
%for each $i_s$ with a hierarchical property (\ref{eq:FiltR}).
As we compute their total conductivities 
and the conductivity curve
$(\sigma_\total($ $\gamma_\agg, p, i_s))_{p \in [0,1]}$ 
of (\ref{eq:1-2}) in the following section, 
it is obvious that the curve 
$(\sigma_\total($ $\gamma_\agg, p, i_s))_{p \in [0,1]}$ 
is
a well-defined function over the path
$(\Rconf_{\gamma_\agg, p, i_s})_{p \in [0,1]}$ 
in a measure space for a sequence of the conditional probabilistic events
\cite{Fel,Kol};
for fixing $\gamma_\agg$,
the curve
$(\sigma_\total($ $\gamma_\agg, p, i_s))_{p \in [0,1]}$ 
corresponds to $(\Rconf_{\gamma_\agg, p, i_s})_{p \in [0,1]}$  for each
$i_s$.
In other words, we can treat the statistical
properties, such as variance and average,
 of the conductivity curves in our method
following the arguments
of the infinite probability fields in \cite[Chap II]{Kol}. 
%}

\begin{figure}[htbp]
 \begin{tabular}{cc}
  \begin{minipage}[t]{0.45\hsize}
   \begin{center}
    \includegraphics[width=\hsize, clip]{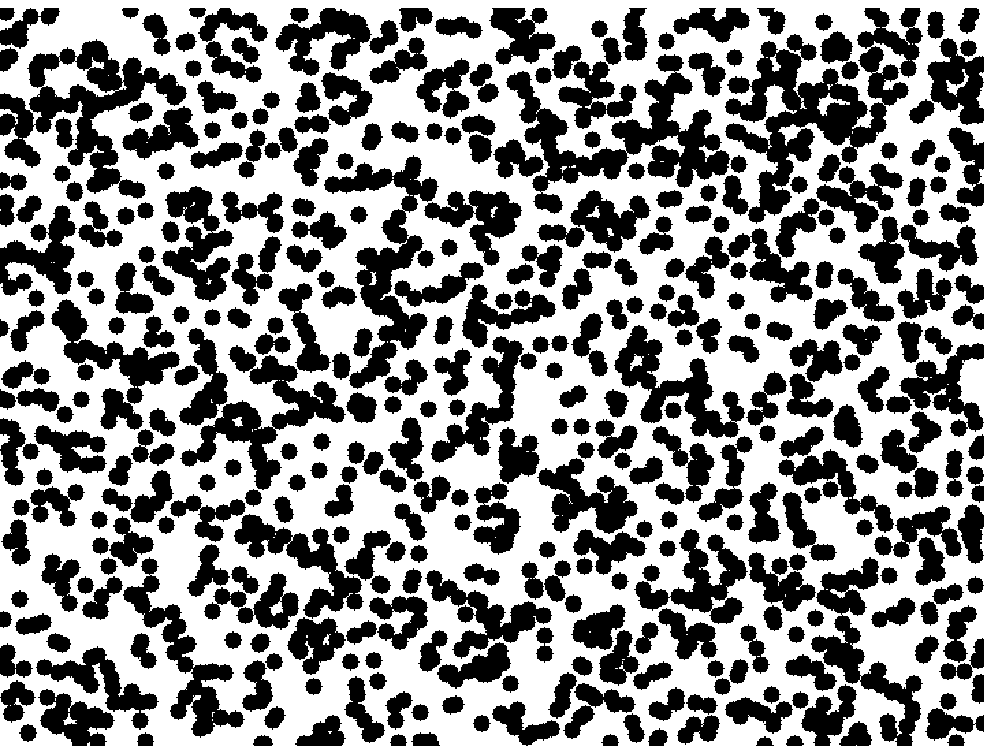}
    (a)
   \end{center}
   \begin{center}
    \includegraphics[width=\hsize, clip]{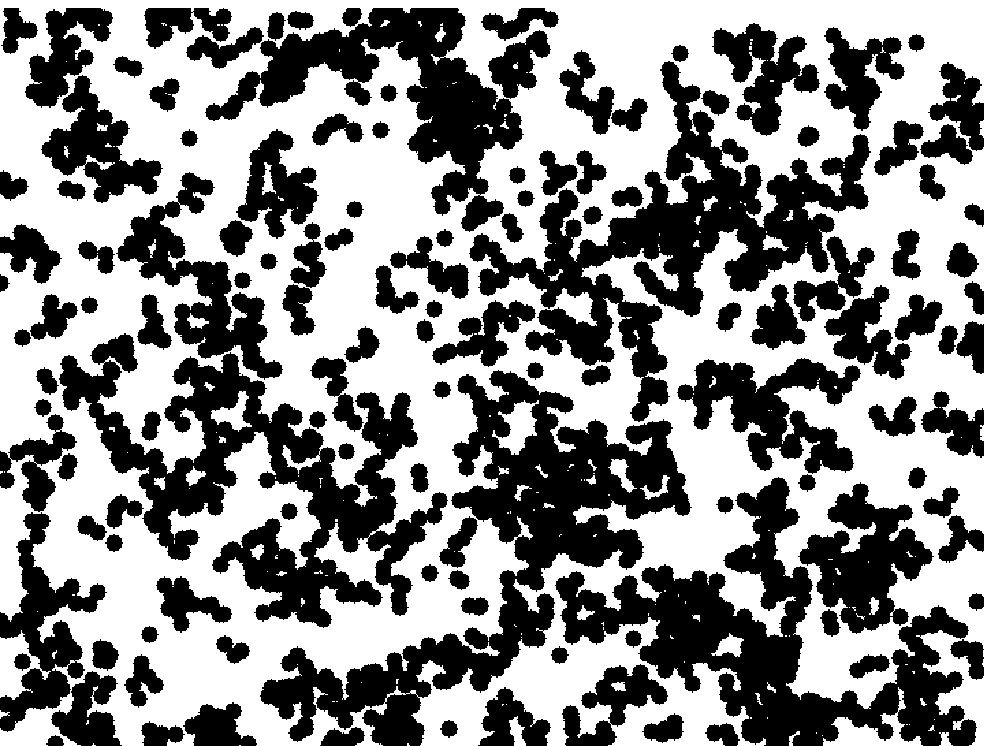}
    (c)
   \end{center}
  \end{minipage} 
  \begin{minipage}[t]{0.45\hsize}
   \begin{center}
    \includegraphics[width=\hsize, clip]{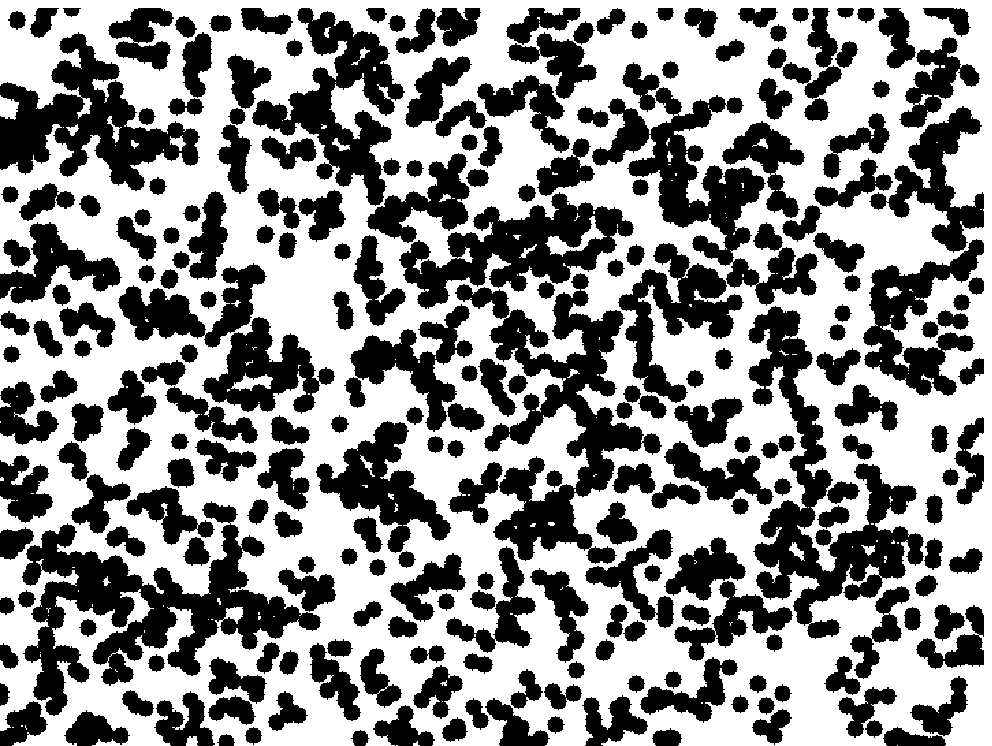}
    (b)
   \end{center}
   \begin{center}
    \includegraphics[width=\hsize, clip]{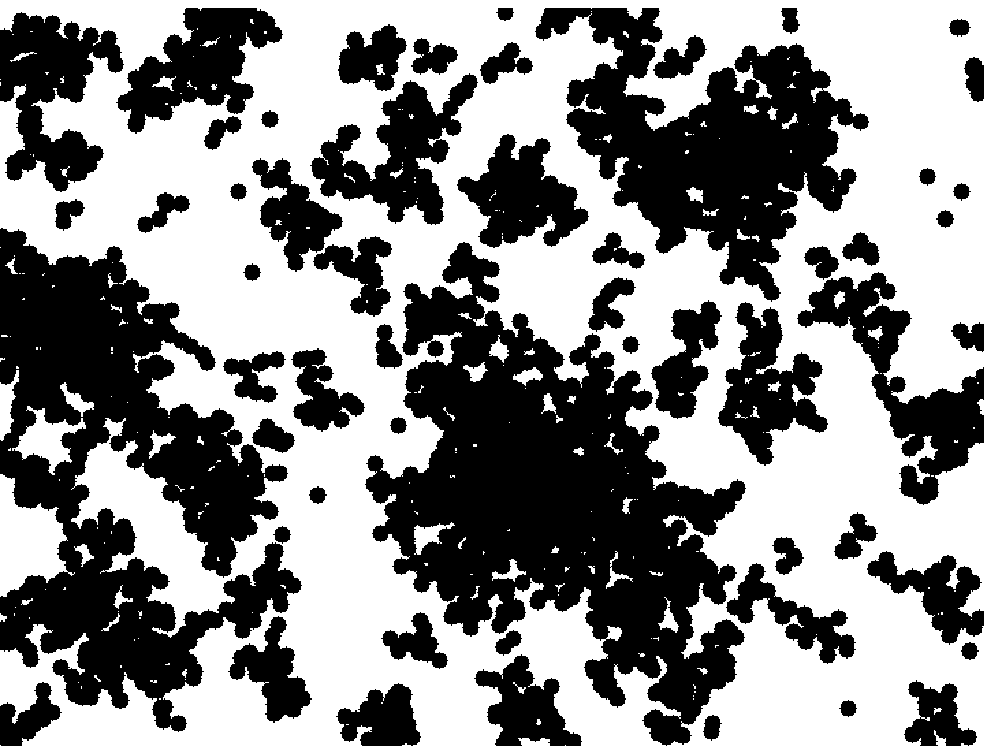}
    (d)
   \end{center}
  \end{minipage} 
 \end{tabular}
\caption{\label{fig:2Dview} The agglomeration parameters:
%\textcolor{blue}{
Two-dimensional
%}
random particle system with the agglomeration parameter 
$\gamma_\agg$
$=0.0, 0.5, 0.8$ and $0.95$ for (a), (b), (c), and (d) respectively.
%\textcolor{blue}{
The radius of the particle is six meshes,
the width and height of the images are 720 and 540 meshes respectively,
and the %\textcolor{red}{\sout{surface-density}}
 area fraction
 of the particles is 0.45.  }
%}
\end{figure}

%\begin{figure}[htbp]
%\begin{center}
%\includegraphics[scale=0.4]{Fig03.eps}
%\end{center}
%\caption{\label{fig:2Dview} The agglomeration parameters:
%Random particle system with the agglomeration parameter 
%$\gamma_\agg$
%$=0.0, 0.5, 0.8$ and $0.95$ for (a), (b), (c) and (d) respectively.}
%\end{figure}

\begin{figure}[htbp]
\begin{center}
\includegraphics[scale=0.35]{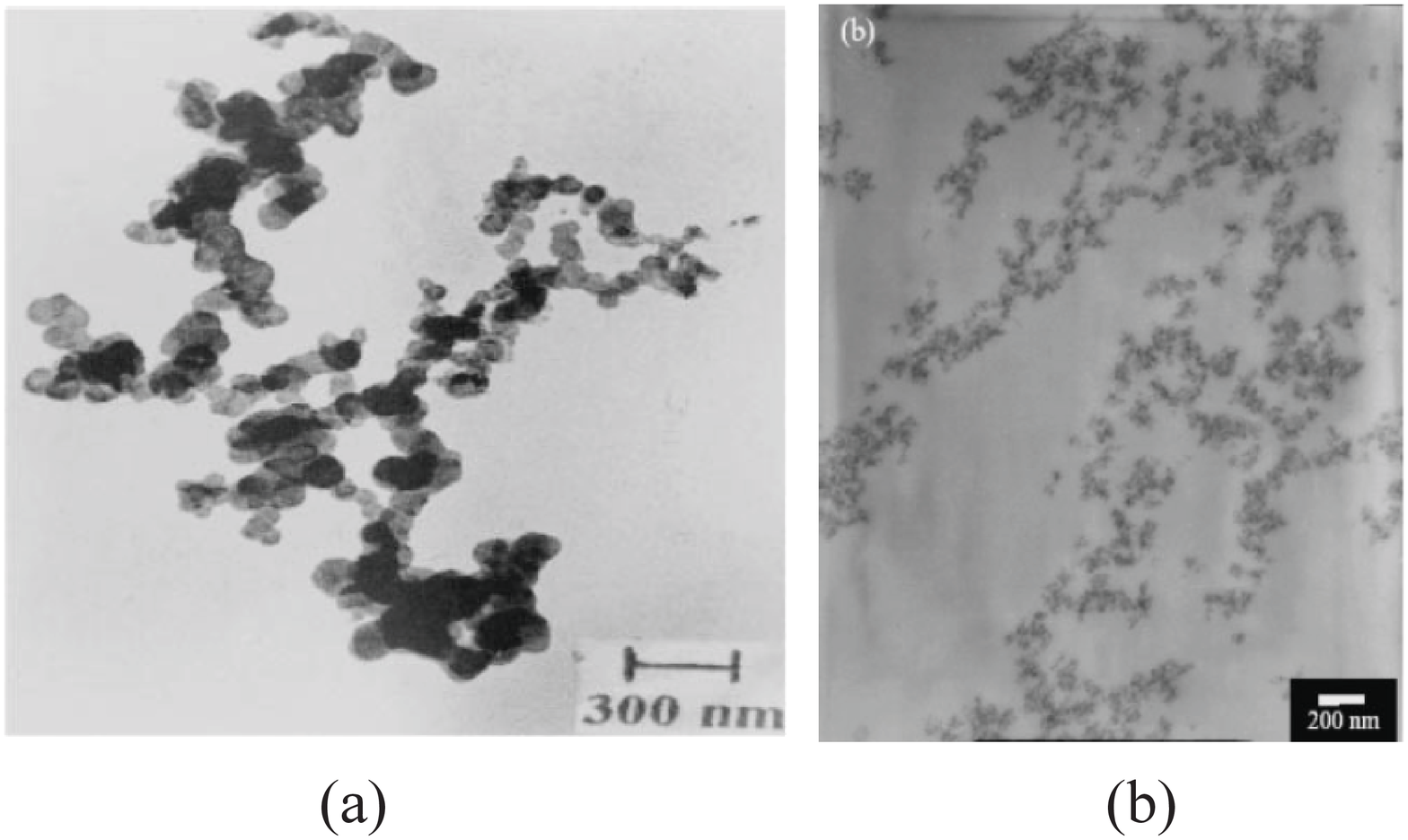}
\end{center}
\caption{\label{fig:TEMimage}
TEM micrograms of 
(a) polyacrylonitrile nanocomposite of acetylene black particles
%in Ref.9 % \cite{MB} 
in Ref.~\cite{MB} 
and
(b) epoxy nanocomposites containing silicate oxide particles
%in Ref.21 % \cite{CLSK} 
in Ref.~\cite{CLSK}. 
}
\end{figure}

In order to illustrate this algorithm, we will show two dimensional case.
In a very similar way to the three-dimensional case, we have two dimensional
agglomeration configurations which are given in Figure \ref{fig:2Dview}.
Figure \ref{fig:TEMimage} exhibits the transmission electron micrograph (TEM)
of the real agglomerated clusters in Refs.~\cite{MB} and \cite{CLSK}.
Ref.~\cite{MB} is of  the composite of polyacrylonitrile
 with acetylene black particles, and Ref.~\cite{CLSK} shows
the epoxy nanocomposites with silicate oxide particles.
These configurations in Figure \ref{fig:2Dview} might simulate some
agglomeration states in Figure \ref{fig:TEMimage}. Even though it is
difficult to visualize the three dimensional cases well, it is expected
that our algorithm generates the similar agglomerate configurations
in Figure \ref{fig:3Dview}.

%\textcolor{blue}{
If we employ other more physical treatments
for agglomeration such as \cite{TTK,KH},
 it is difficult 
to find a parameter which directly represents the agglomeration,
and is also hard
to have a natural   hierarchical property such as (\ref{eq:FiltR}).
In other words, we cannot basically define the conductivity curve
as a function over a path in a measure space though
of course, 
a single conductivity curve over $p \in [0,1]$
can be defined as a fitting curve
for total conductivities to configurations. 
It means that our method has an advantage as a mathematical model.
%}

\subsection{Computation of the conductivity in ACPMs}

To apply FDM \cite{LV} to the computation of the conductivity in ACPM,
we use a $N_x \times N_y \times N_z$ lattice denoted by $\cL$
to represent the box-region $\cB$ by the integers $N_x$, $N_y$ and $N_z$. 
In this article, we mainly assume that $N_x = N_y = N_z = 216$ and
the radius of the particle $\rho = 1$ corresponds to six meshes. 
It means that the ratio between the volumes of $\cB$ and a particle is
about $1.1 \times 10^4$ and the size of 
$\cB = [0,L]^3$ is given as $L \approx 36$.

Further we set the conductivity distribution $\sigma(x, y, z)$
which consists of the conductive particles and the insulator as a background
in the box-region $\cB$ as in Figure \ref{fig:3Dview}.
We put the local conductivity $\sigma_\mat = 1$ inside of each
$\Rconf_{\gamma_\agg, p,i_s}$.  As the insulator, we set the infinitesimal
conductivity $\sigma_{\inf}=10^{-4}$ within background.
In other words, we handle binary materials with largely different
local conductivities $\sigma_\mat$ and $\sigma_\inf$.

In order to compute the total conductivity, we set the voltage
$\phi=\phi_0 = 1$ and $\phi=0$ on the upper and the lower faces, {\it{i.e.}},
$[0,x_0]\times [0,y_0]\times \{z_0\}$ and
$[0,x_0]\times [0,y_0]\times \{0\}$
respectively as the boundary condition corresponding to the electrodes.
As the side boundary condition, we used the natural boundary for each
$yz$-face and each $xz$-face.  In other words, at the side boundaries,
we imposed that current normal to each face vanished.

Following our algorithm of FDM as  mentioned in detail in Ref.~\cite{MSW},
we numerically solved the generalized Laplace equation over the region,
\begin{equation}
	\nabla \cdot \sigma \nabla \phi =0,
\label{eq:1-1}
\end{equation}
for the conductivity distribution $\sigma(x,y,z)$ for each 
$\Rconf_{\gamma_\agg, p, i_s}$ to obtain the potential
distribution $\phi(x,y,z)$.
By numerically solving (\ref{eq:1-1}), we obtained the total conductivity
$\sigma_\total$ of the system after we integrated the current
$\sigma \nabla \phi$ over a $xy$-plane.
Since $\sigma_\total$ is determined for each $\Rconf_{\gamma_\agg, p,i_s}$,
$\sigma_\total$ is a function of the volume fraction $p$, 
%\textcolor{blue}{
the agglomeration parameter
%} 
$\gamma_\agg$ and the seed $i_s$ of the pseudo-randomness. We denote it by 
$\sigma_\total(p, \gamma_\agg, i_s)$ explicitly or simply $\sigma_\total(p)$.

As mentioned in Introduction, the conductivity curve 
$\sigma_\total(p)$  $(p \in [0,1])$ is expressed by
\begin{equation}
\sigma_\total(p) = 
    \left\{ \begin{array}{ll}
         \frac{(p - p_c)^\talpha}{(1.0 - p_c)^\talpha}, & \mbox{for} \ 
            p \in [p_c, 1],\\
          0 & \mbox{otherwise},\\
      \end{array} \right.
\label{eq:sigmatotal}
\end{equation}
where $p_c$ is the threshold and $\talpha$ is the critical exponent,
or merely exponent.  
%\textcolor{blue}{
As mentioned in Subsection 2.1, 
the conductivity curve is a function over a path
$(\Rconf_{\gamma_\agg, p,i_s})_{p \in [0,1]}$ in the
measure space, which is characterized by 
the agglomeration parameter $\gamma_\agg$
and the seed $i_s$ of the pseudo-randomness.
Thus 
 the threshold  $p_c$ and the exponent
$\talpha$ are naturally determined for the individual curve and
can be regarded as functions of 
the agglomeration parameter $\gamma_\agg$ and the seed $i_s$,
which we sometimes express as
$p_c(\gamma_\agg, i_s)$ and $\talpha(\gamma_\agg, i_s)$.
%}
%\textcolor{red}{\sout{
%Since the threshold  $p_c$ and the exponent
%$\talpha$ are determined by
%$\{\Rconf_{\gamma_\agg, p,i_s} \ |$ $ \ p \in [0,1]\}$,
% they are functions
%of the agglomeration parameter $\gamma_\agg$ and the seed $i_s$.
%Thus we sometimes express them as $p_c(\gamma_\agg, i_s)$ and
%$\talpha(\gamma_\agg, i_s)$.
%}}

%\textcolor{blue}{
Therefore
%}
we evaluated the threshold $p_c(\gamma_\agg, i_s)$ and
the exponent $\talpha(\gamma_\agg, i_s)$ as
the fitting parameters so that each average of the square error from the 
curve is the smallest, by monitoring the square root of the average
of the square error.  The square root of the average of the square error
is denoted by $\delta \sigma_\total$
$ \equiv \delta \sigma_\total(\gamma_\agg, i_s)$.

\section{Results}

For each seed $i_s$ of the pseudo-randomness,
%\textcolor{red}{\sout{
%a configuration $\Rconf_{\gamma_\agg, p,i_s}$ to a volume fraction $p$
%naturally contains the configuration $\Rconf_{\gamma_\agg, p',i_s}$
%to $p'<p$ of the same seed $i_s$.  Hence for a given seed $i_s$, 
%the total conductivities $\sigma_\total$ between
%$\Rconf_{\gamma_\agg, p,i_s}$ and $\Rconf_{\gamma_\agg, p',i_s}$ are
%}}
%\textcolor{blue}{
the total conductivities $\sigma_\total(p,$ $\gamma_\agg, i_s)$ 
are obtained over $\{\Rconf_{\gamma_\agg, p,i_s} \ | \ p\in [0,1]\}$
with the hierarchical structure (\ref{eq:FiltR})
as a conditional probabilistic problem.
In other words, for each $i_s$,
%}
%\textcolor{blue}{
%} 
 we obtained the conductivity curve
$(\sigma_\total(p,\gamma_\agg, i_s))_{p\in[0,1]}$ 
over the $(\Rconf_{\gamma_\agg, p,i_s})_{p\in [0,1]}$ as a
 %\textcolor{red}{\sout{smooth curve}}
%\textcolor{blue}{
measurable function
%}
 even with the non-vanishing $\gamma_\agg$;
%\textcolor{blue}{
%as a function over the path of the conditional probabilistic problem; 
if need be, we can justify it using the arguments
of the infinite probability fields in \cite[Chap II]{Kol}.
%It corresponds to the conductivity of a sample of small
%devices with nanoparticle.
%} 

\begin{figure}[htbp]
 \begin{tabular}{cc}
  \begin{minipage}[t]{0.5\hsize}
   \begin{center}
    \includegraphics[width=\hsize, clip]{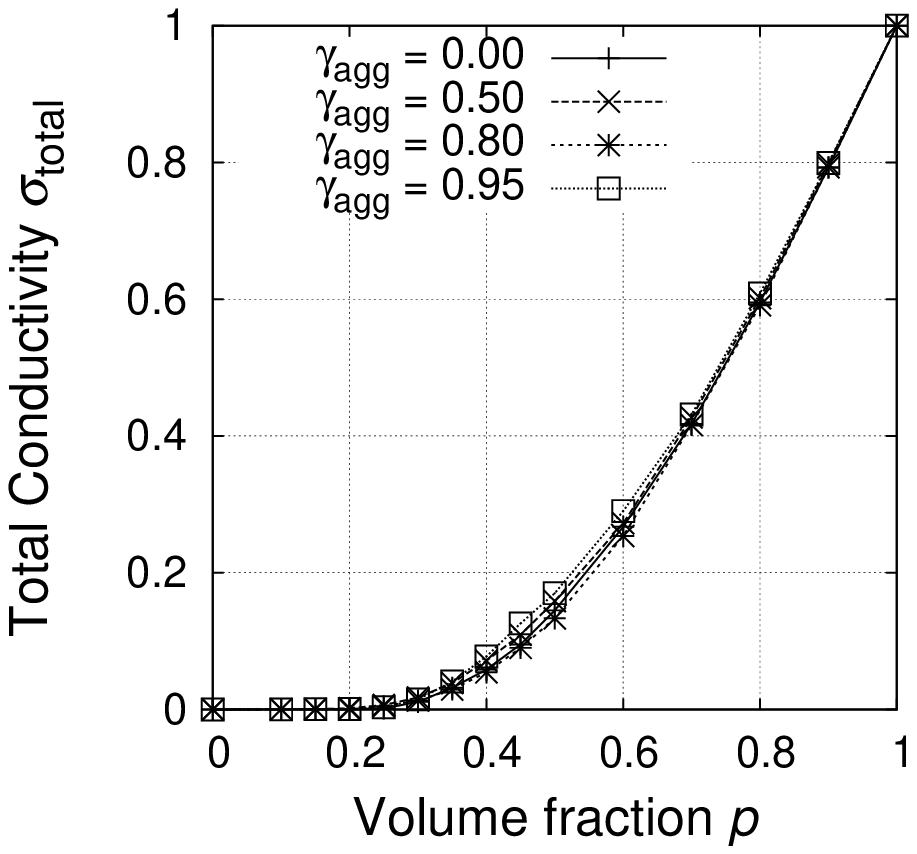}
    (a)
   \end{center}
  \end{minipage} 
  \begin{minipage}[t]{0.5\hsize}
   \begin{center}
    \includegraphics[width=\hsize, clip]{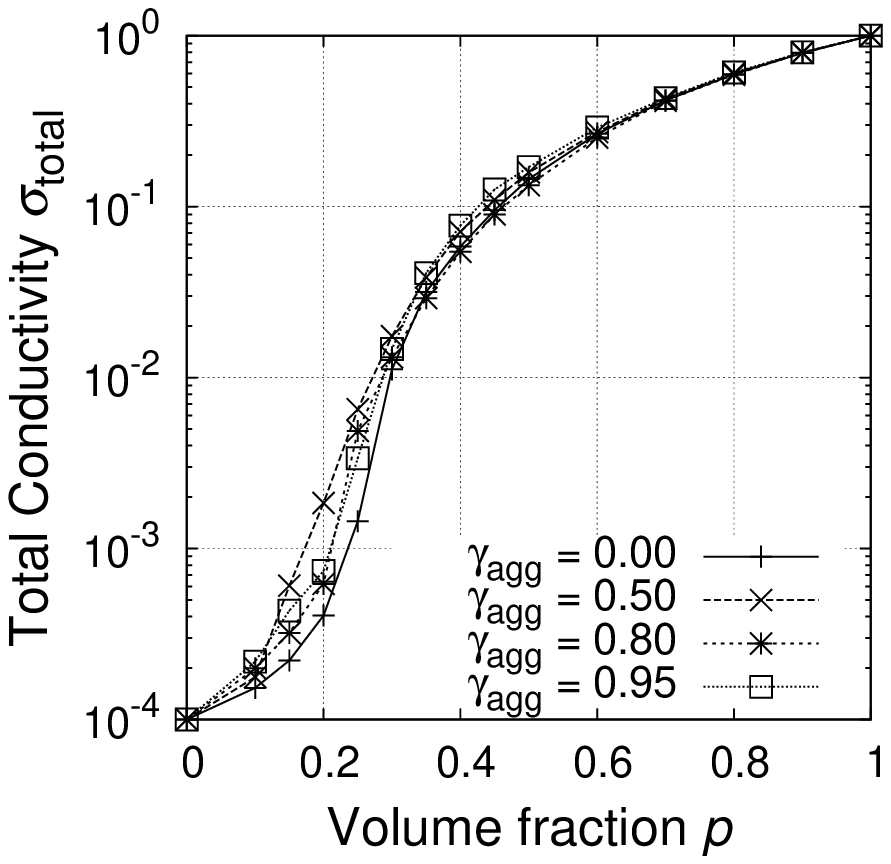}
    (b)
   \end{center}
  \end{minipage} 
 \end{tabular}
\caption{\label{fig:pcurve} The conductivity curves for
agglomeration parameters $\gamma_\agg = 0.0, 0.5, 0.8$ and
$0.95$ of 
%\textcolor{blue}{
the seed $i_s = 1$ of the pseudo-randomness:
%}
(a) is of the linear scale
and (b) is of the logarithm scale.}
\end{figure}

\begin{figure}[htbp]
 \begin{tabular}{ccc}
  \begin{minipage}[t]{0.33\hsize}
   \begin{center}
    \includegraphics[width=\hsize, clip]{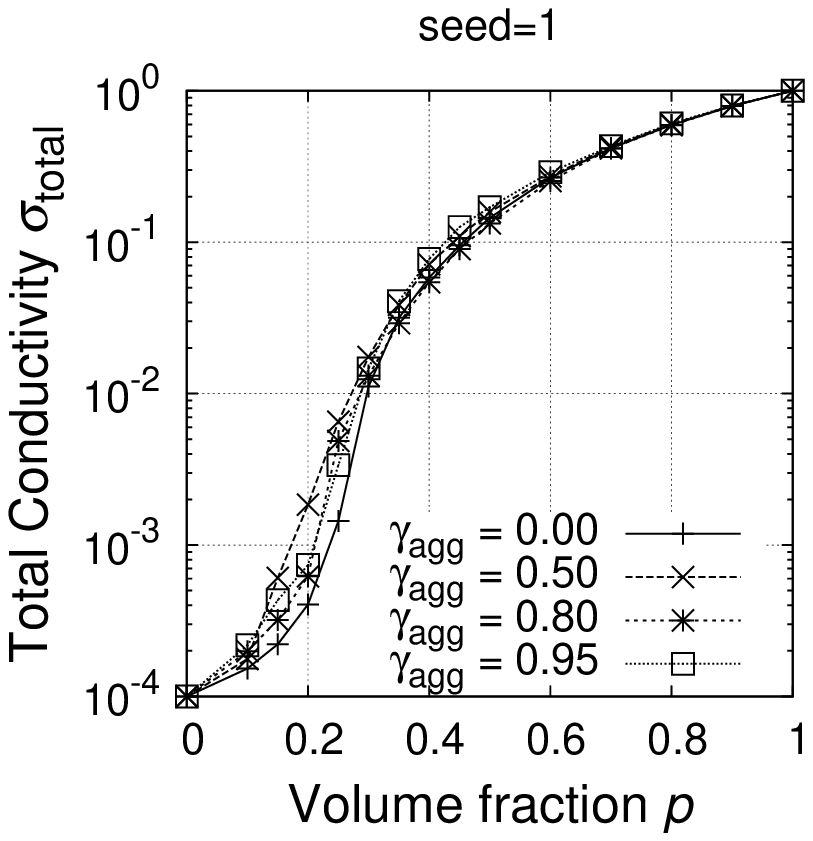}
    \includegraphics[width=\hsize, clip]{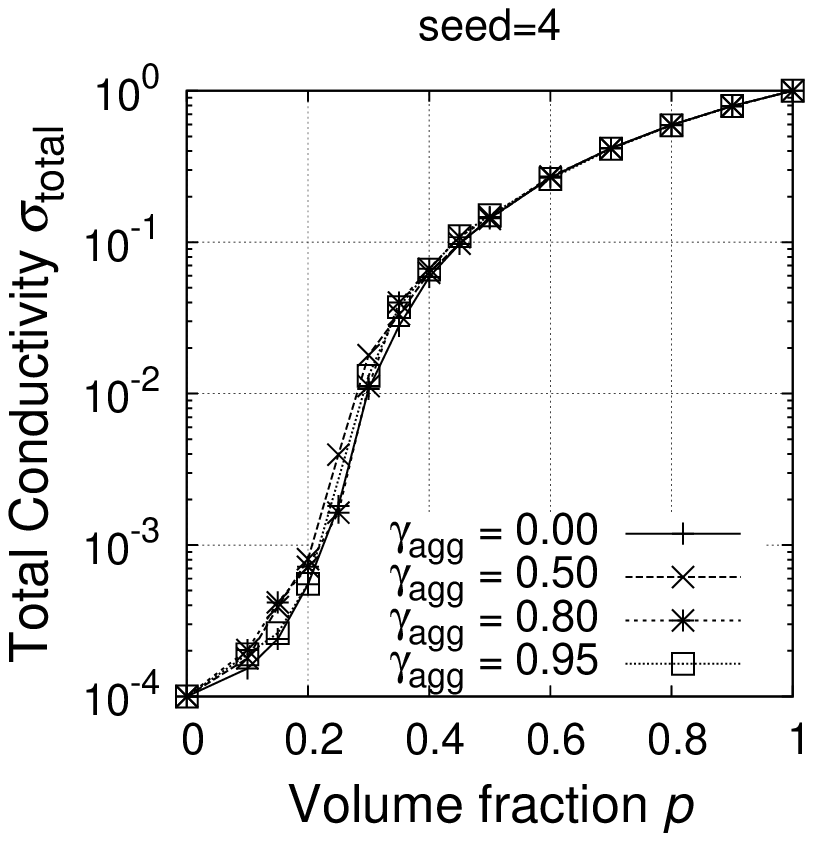}
    \includegraphics[width=\hsize, clip]{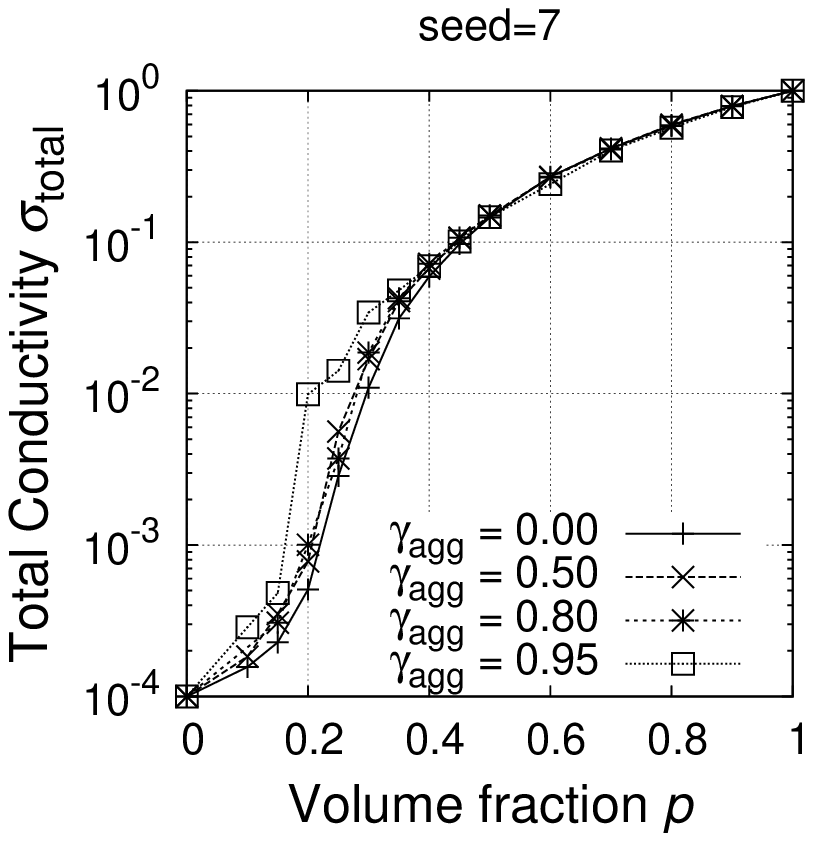}
    \includegraphics[width=\hsize, clip]{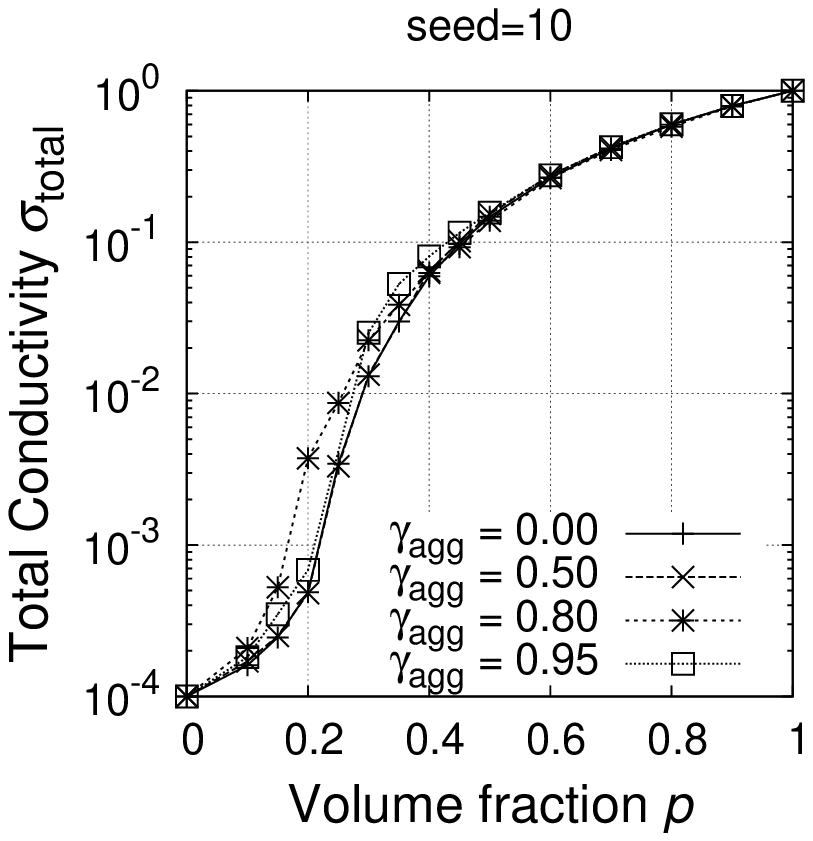}
   \end{center}
  \end{minipage} 
  \begin{minipage}[t]{0.33\hsize}
   \begin{center}
    \includegraphics[width=\hsize, clip]{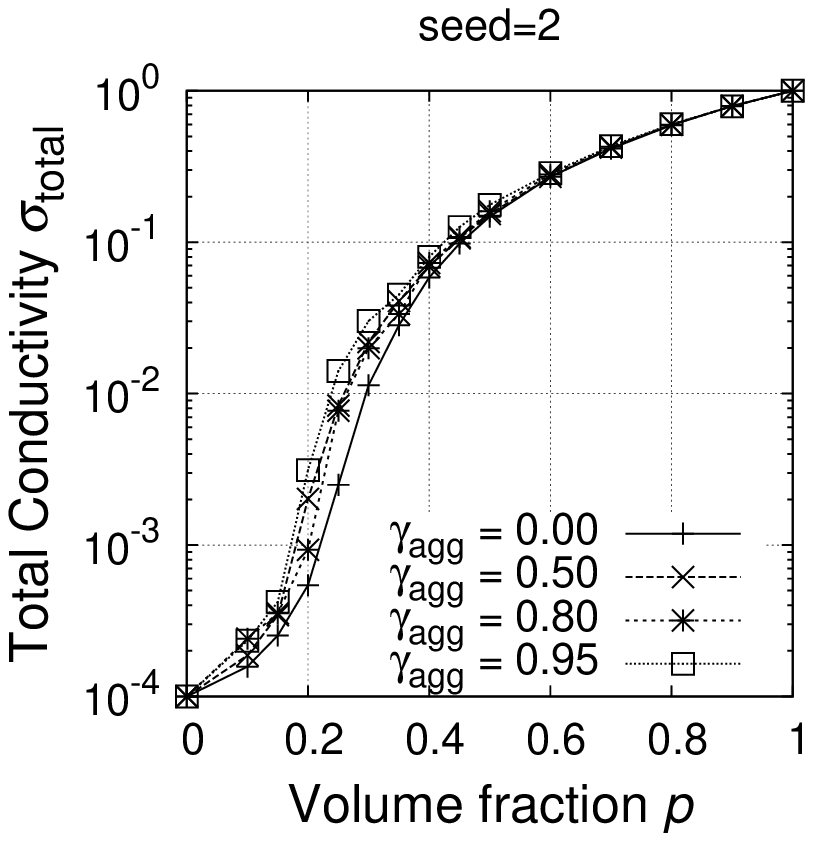}
    \includegraphics[width=\hsize, clip]{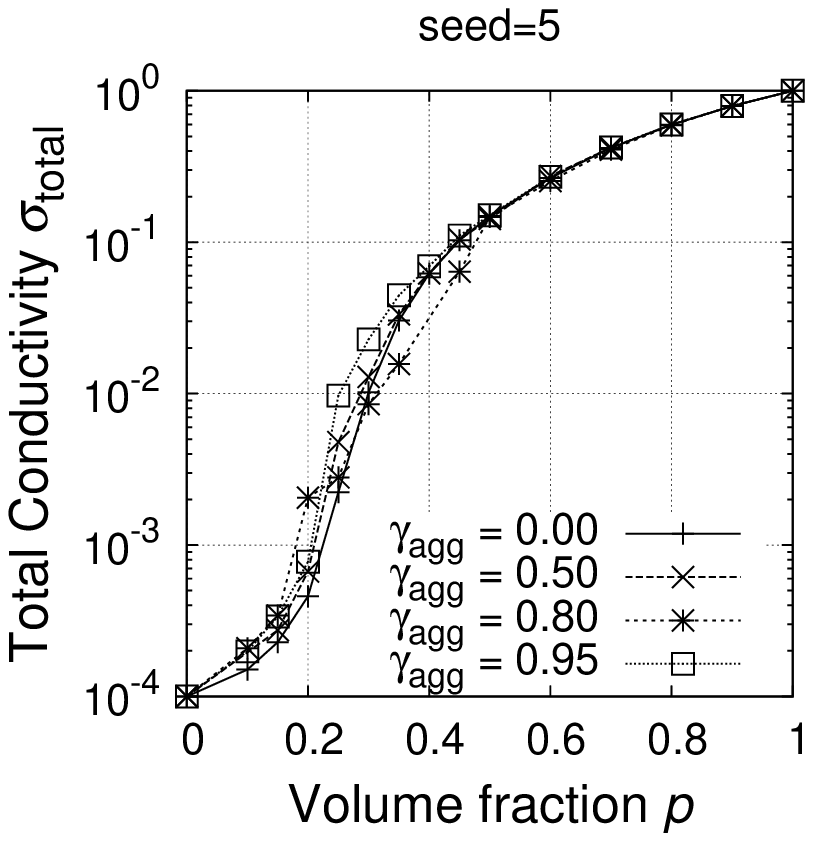}
    \includegraphics[width=\hsize, clip]{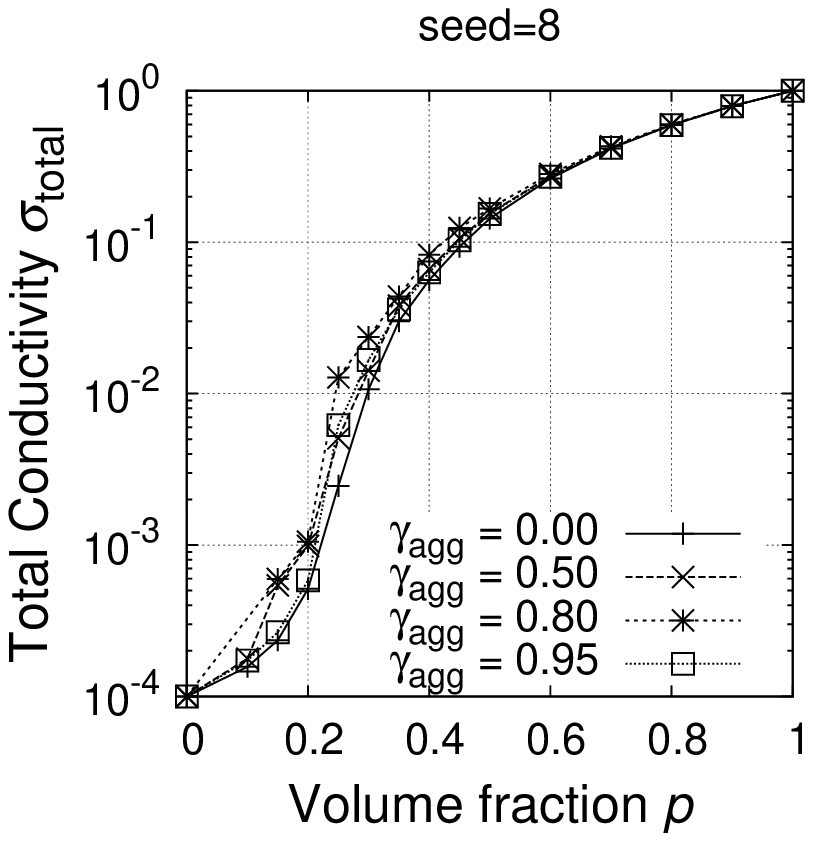}
    \includegraphics[width=\hsize, clip]{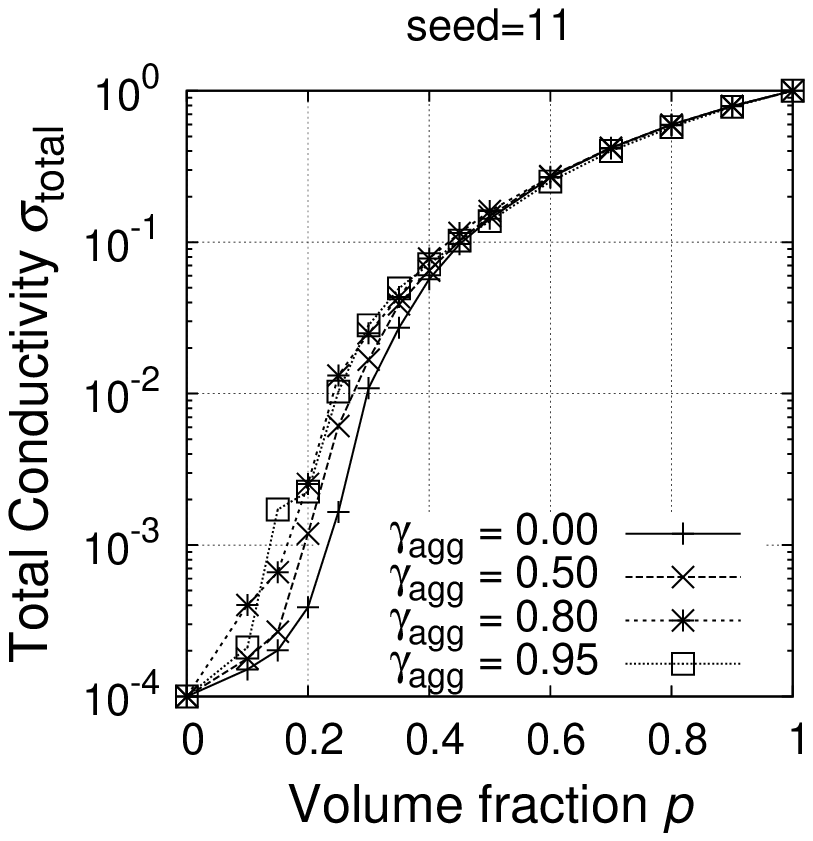}
   \end{center}
  \end{minipage} 
  \begin{minipage}[t]{0.33\hsize}
   \begin{center}
    \includegraphics[width=\hsize, clip]{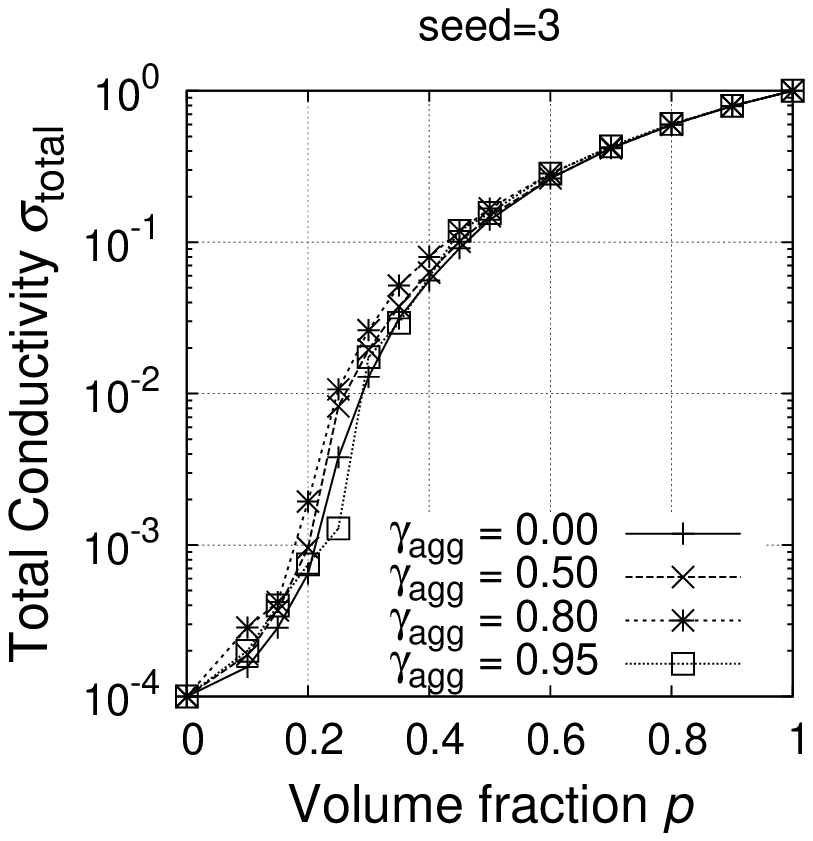}
    \includegraphics[width=\hsize, clip]{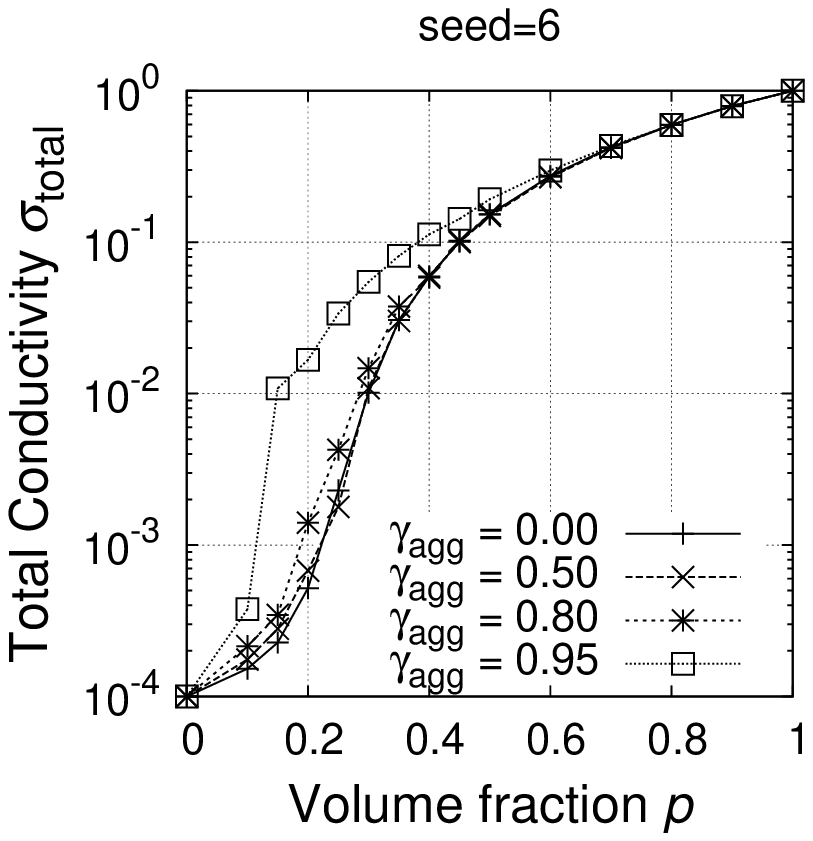}
    \includegraphics[width=\hsize, clip]{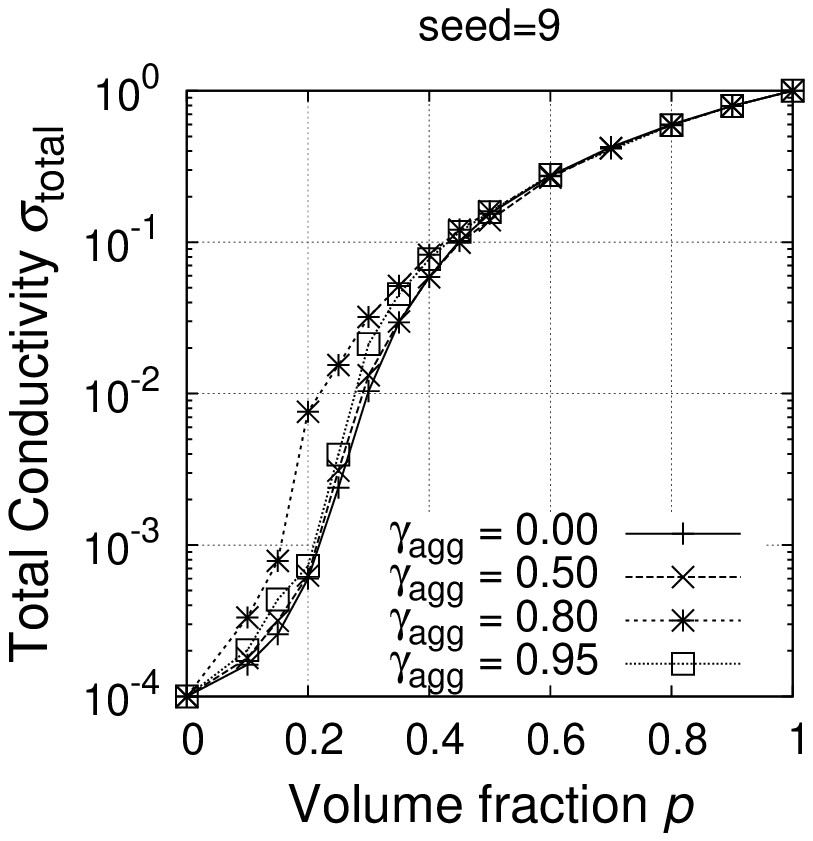}
    \includegraphics[width=\hsize, clip]{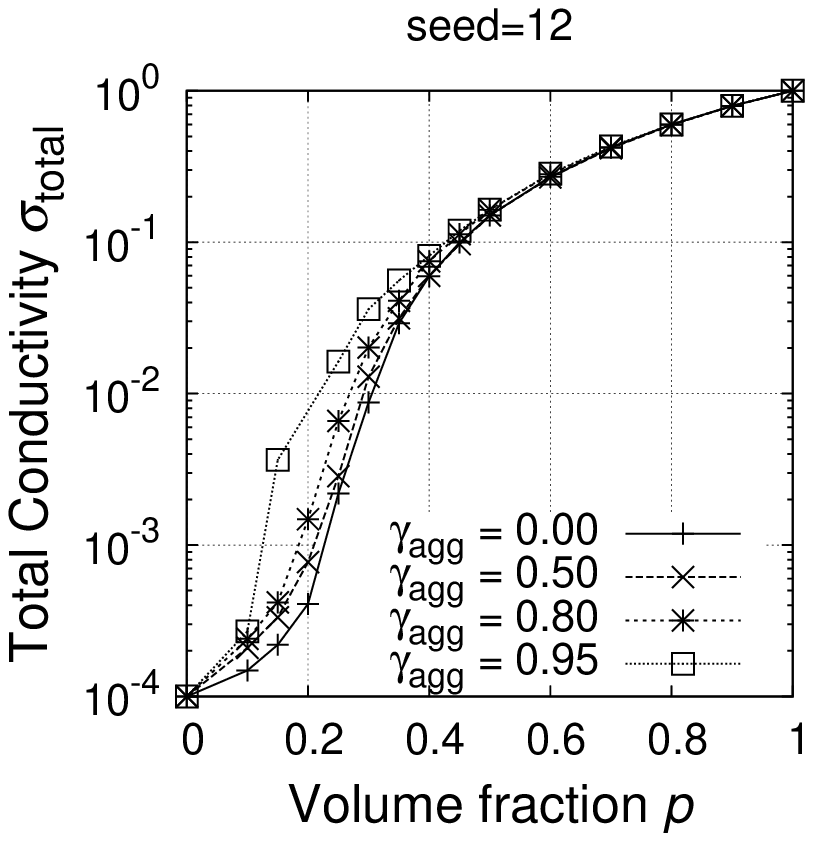}
   \end{center}
  \end{minipage} 
 \end{tabular}
\caption{\label{fig:pcurve_allseeds_1} The conductivity curves for
agglomeration parameters $\gamma_\agg = 0.0, 0.5, 0.8$ and
$0.95$ of the 
%\textcolor{blue}{
seed $i_s=1, 2, \dots, 12$ of the pseudo-randomness.
%}
}
\end{figure}

\begin{figure}[htbp]
 \begin{tabular}{ccc}
  \begin{minipage}[t]{0.33\hsize}
   \begin{center}
    \includegraphics[width=\hsize, clip]{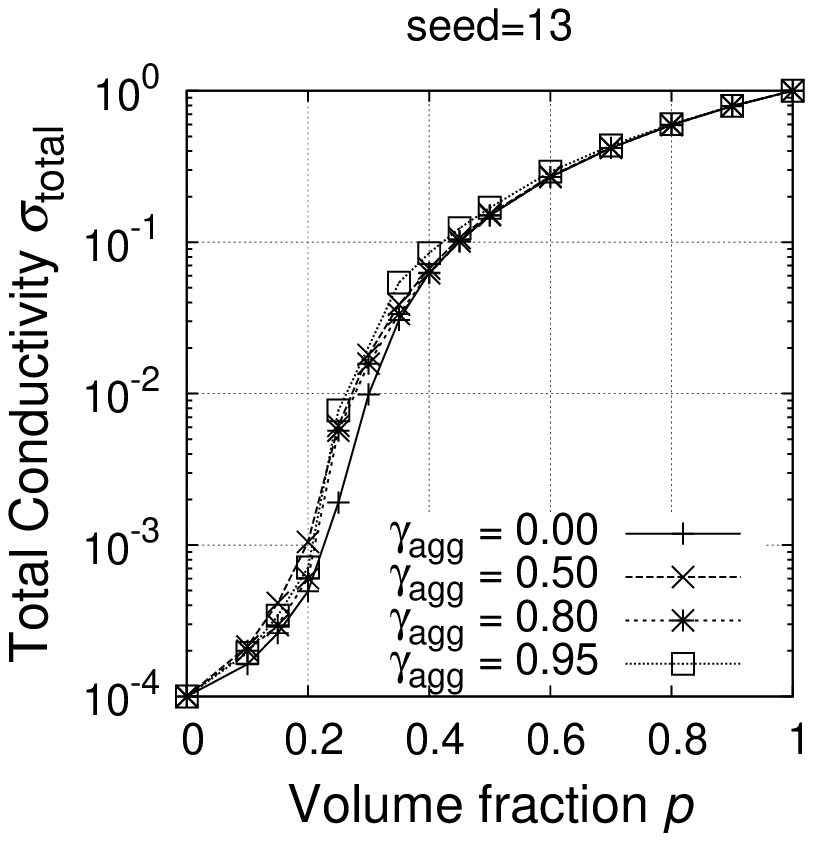}
    \includegraphics[width=\hsize, clip]{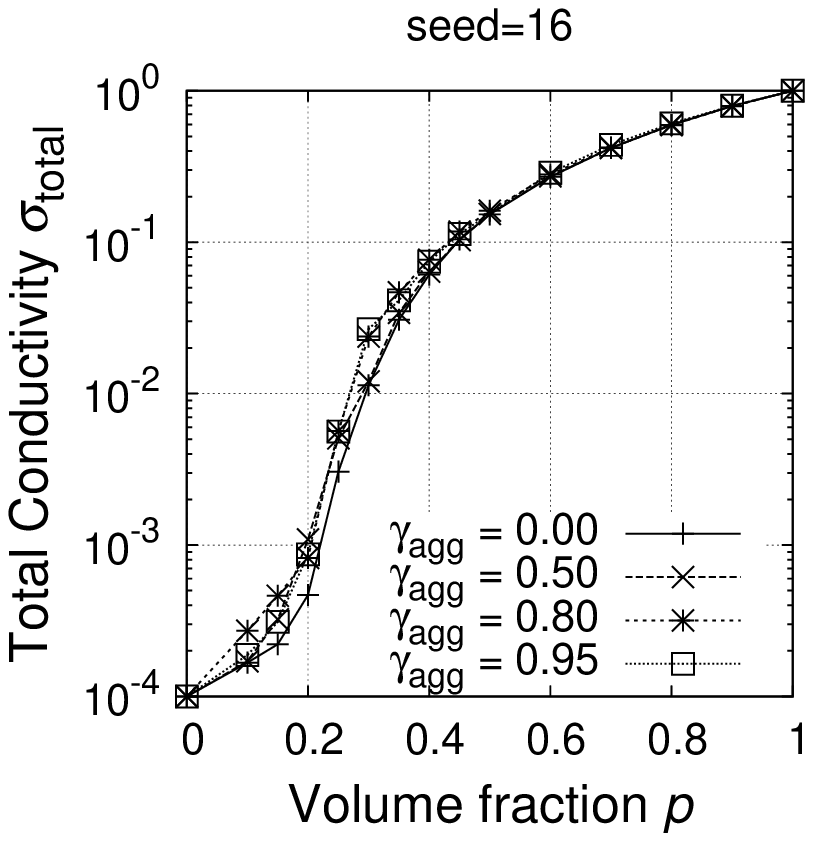}
    \includegraphics[width=\hsize, clip]{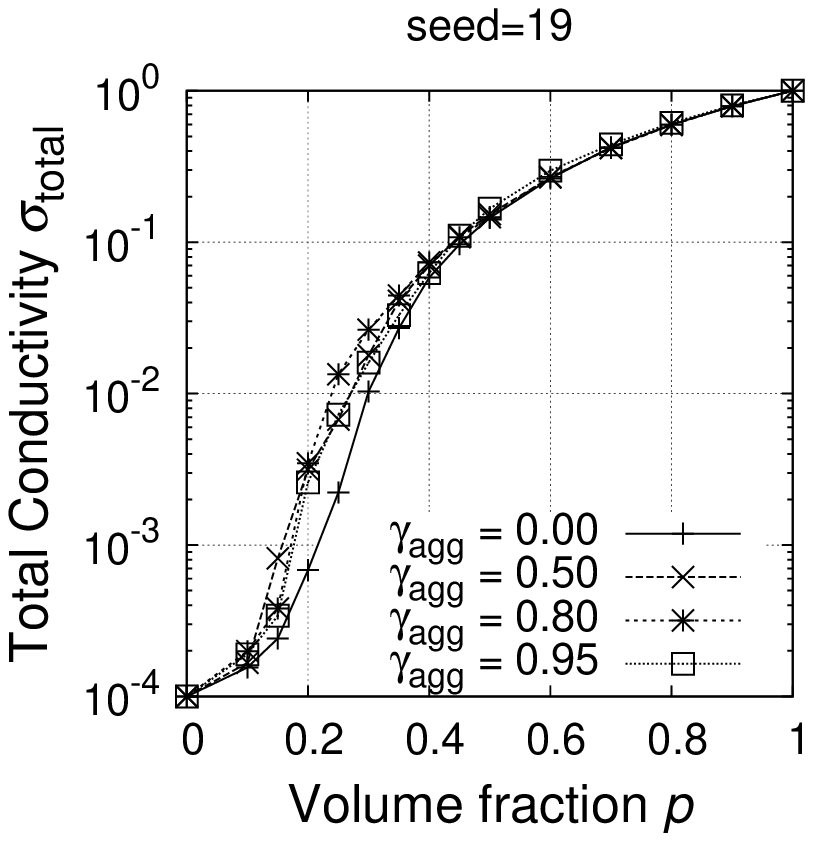}
   \end{center}
  \end{minipage} 
  \begin{minipage}[t]{0.33\hsize}
   \begin{center}
    \includegraphics[width=\hsize, clip]{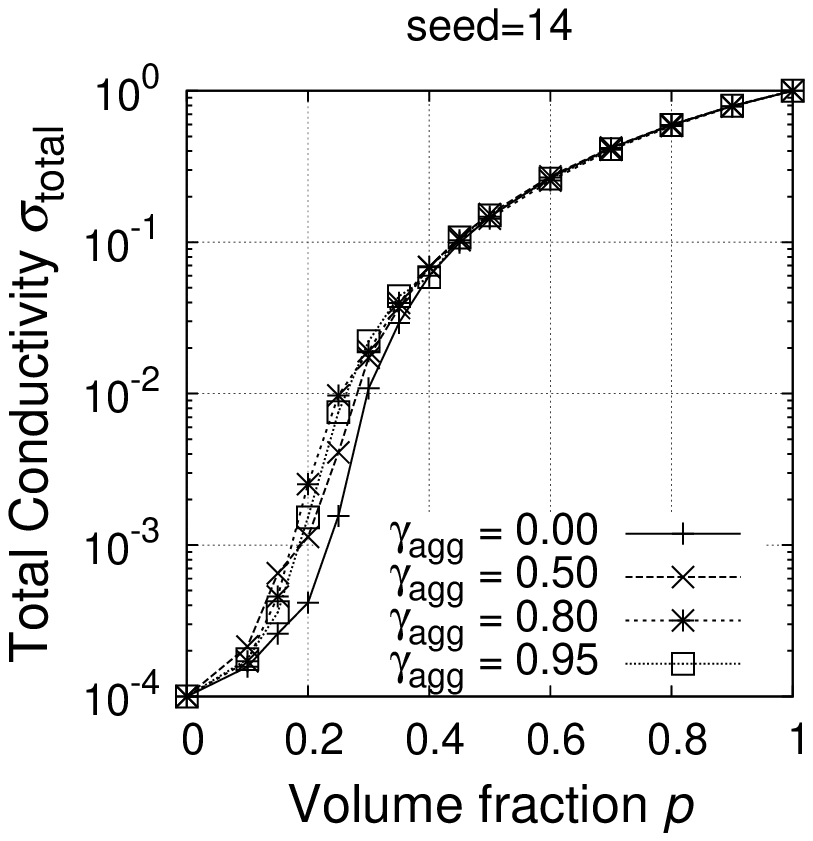}
    \includegraphics[width=\hsize, clip]{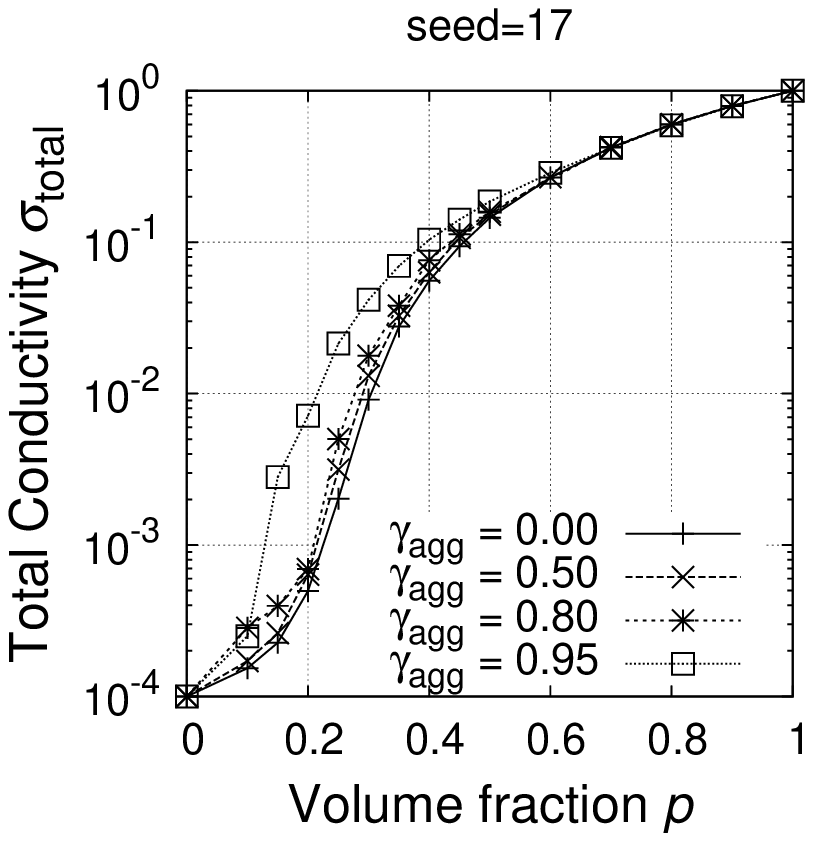}
    \includegraphics[width=\hsize, clip]{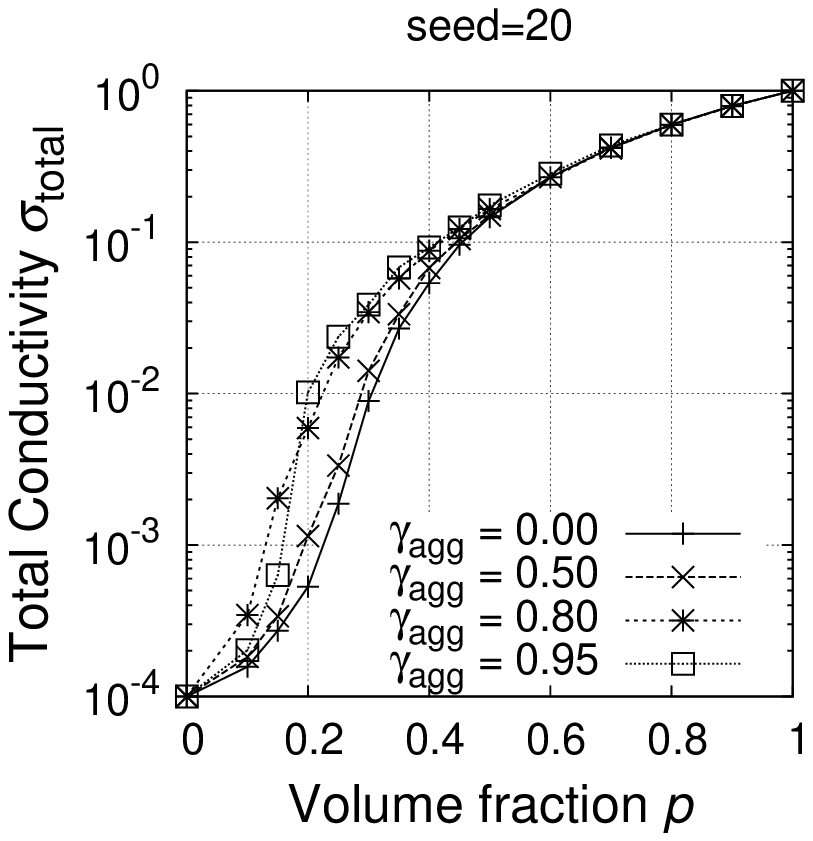}
   \end{center}
  \end{minipage} 
  \begin{minipage}[t]{0.33\hsize}
   \begin{center}
    \includegraphics[width=\hsize, clip]{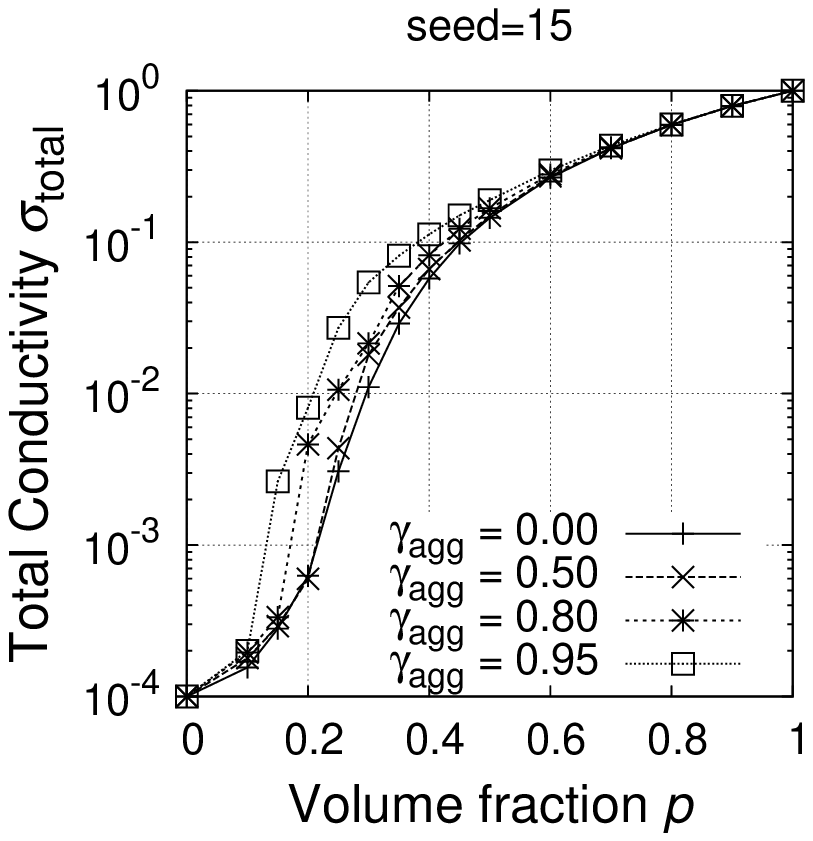}
    \includegraphics[width=\hsize, clip]{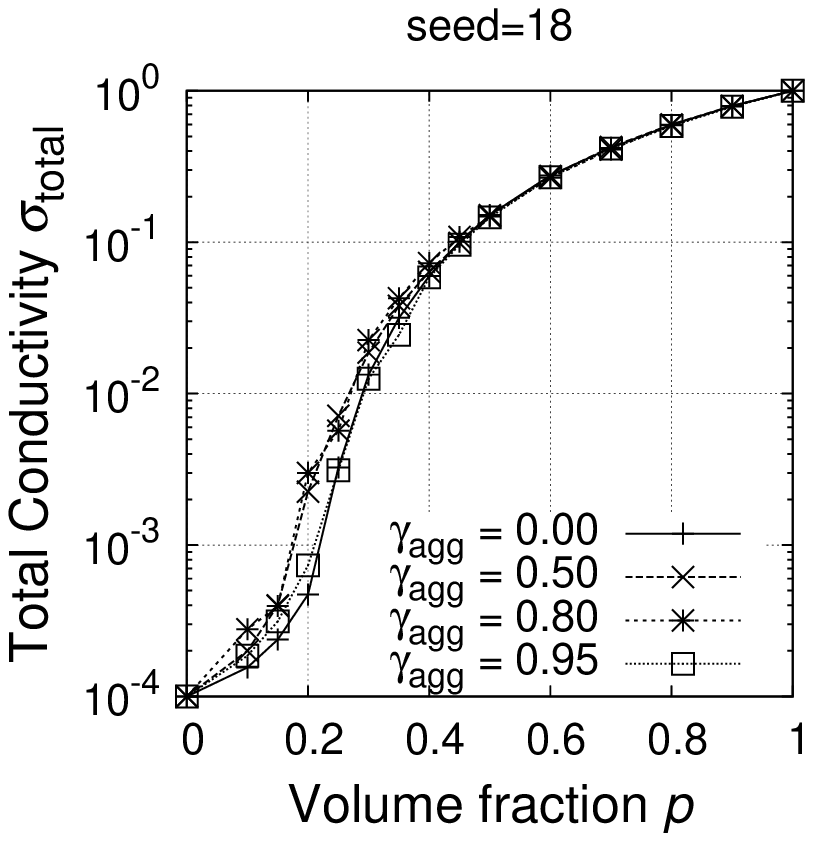}
   \end{center}
  \end{minipage} 
 \end{tabular}
\caption{\label{fig:pcurve_allseeds_2} The conductivity curves for
agglomeration parameters $\gamma_\agg = 0.0, 0.5, 0.8$ and
$0.95$ of the 
%\textcolor{blue}{
seed $i_s=13, 14, \dots, 20$ of the pseudo-randomness.
%}
}
\end{figure}

Figure \ref{fig:pcurve} illustrates the conductivity curves for 
%\textcolor{blue}{
the seed $i_s=1$ of the pseudo-randomness.
%}
Figure \ref{fig:pcurve}(b) shows that we handled the binary 
conductive materials.
%\textcolor{blue}{
The dependence of agglomeration parameters on 
the conductivity curves for 
seed $i_s$ of the pseudo-randomness
in Figures \ref{fig:pcurve_allseeds_1} and \ref{fig:pcurve_allseeds_2}.
%}

\begin{figure}[htbp]
 \begin{center}
  \includegraphics[width=0.7\hsize, clip]{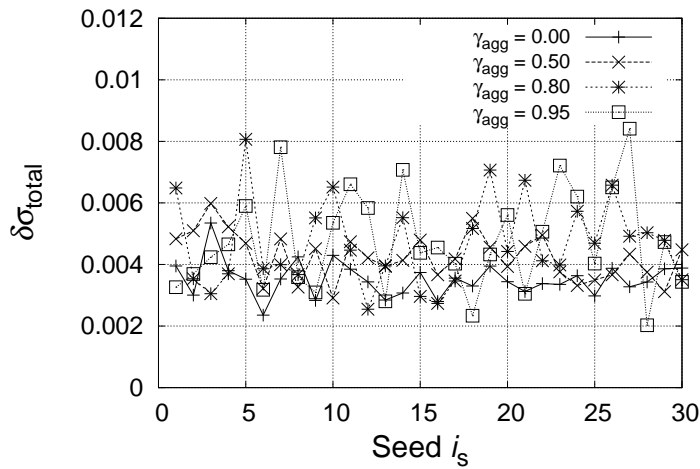}
 \end{center}
\caption{\label{fig:fittingerr}
The fitting errors 
$\delta \sigma_\total$ v.s.~the 
%\textcolor{blue}{
seed $i_s$ of the pseudo-randomness
%}
for $\gamma_\agg = 0.0, 0.5, 0.8$ and $0.95$.
}
\end{figure}

When $p = 0$, the total conductivity $\sigma_\total$ is equal to 
$\sigma_\inf$ which stands for the insulator and is negligible
if we dealt with $\sigma_\total$ linearly as shown in 
Figure \ref{fig:pcurve}(a).
By using the least square error fitting method, we evaluated
the threshold $p_c(\gamma_\agg, i_s)$ 
and the exponent $\talpha(\gamma_\agg, i_s)$ from the curve.
In the evaluation, we used the linear scale of $\sigma_\total$
or the curves in Figure \ref{fig:pcurve}(a).
Figure \ref{fig:fittingerr} shows  the fitting errors 
$\delta \sigma_\total$ v.s.~the 
%$\textcolor{blue}{
seed $i_s$ of the pseudo-randomness;
%}
$\delta \sigma_\total$ is the average of the squares of the 
deviations from the curve (\ref{eq:sigmatotal}) over the 
fitting points for each $\gamma_\agg$ and $i_s$; the fitting 
points are $0.1$, $0.15$, $0.2$, $0.25$, $0.3$, $0.35$,
$0.4$, $0.45$, $0.5$, $0.6$, $0.7$, $0.8$, and $0.9$.
We computed thirty curves for each case, $\gamma_\agg = 0.0$, 
$0.5$, $0.8$ and $0.95$.
The deviations $\delta \sigma_\total$ are not large even for
non-vanishing $\gamma_\agg$ compared with $\sigma_\mat = 1$.
The distribution of $\delta \sigma_\total$ shows that 
the larger $\gamma_\agg$ is, the larger the values are.
Its maximum is 
$\delta \sigma_{\total, \mathrm{max}} \le 0.0084$
for $\gamma_\agg = 0.95$, but the average
$\overline{\delta \sigma_\total}$ is less than 
0.0048;
the values of $(\gamma_\agg, \overline{\delta \sigma_\total})$ are 
given as 
$(0.0, 0.0035)$, $(0.5, 0.0043)$, $(0.8, 0.0047)$ and $(0.95, 0.0048)$.

It means that  the thresholds $p_c$ and the exponents $\talpha$
represent the conductivity curves well in our computations.
In fact since we have
$$
\frac{\partial \sigma_\total(p)}{\partial p_c}
  =-\sigma_\total(p)\frac{1-p}{(1-p_c)(p-p_c)} t, \quad
\frac{\partial \sigma_\total(p)}{\partial t}
  =\sigma_\total(p)\log\frac{p-p_c}{1-p_c},
$$
the accuracies $\delta_\sigma p$ and $\delta_\sigma t$ of $p_c$ and $t$,
 could be estimated using $\delta \sigma_\total$;
$$
\delta_\sigma p = 
\left|\left(\frac{\partial \sigma_\total}{\partial p_c}\right)^{-1} 
\right|\overline{\delta \sigma_\total},
\quad
\delta_\sigma t = \left|
\left(\frac{\partial \sigma_\total}{\partial t}\right)^{-1}
\right| \overline{\delta \sigma_\total}.
$$
For example, around $p \approx 0.5$ $(\sigma_\total \approx 0.2)$, 
they are evaluated as 
%%SHIM_changed
%\befor{$\delta_\sigma p \approx \overline{\delta\sigma_\total} \approx 0.002$}
%\after{$\delta_\sigma p \approx \overline{\delta\sigma_\total} \approx 0.005$}
$\delta_\sigma p \approx \overline{\delta\sigma_\total} \approx 0.005$
%%SHIM_changed
and 
%%SHIM_changed
%\befor{$\delta_\sigma t \approx 5 \overline{\delta \sigma_\total} \approx 0.01$}
%\after{$\delta_\sigma t \approx 4.3 \overline{\delta \sigma_\total} \approx 0.022$}
$\delta_\sigma t \approx 4.3 \overline{\delta \sigma_\total} \approx 0.022$
%%SHIM_changed
by setting
%%SHIM_changed
%\befor{$p_c \sim 0.25$, $t\sim 1.8$ as shown later.}
%\after{$p_c \sim 0.27$, $t\sim 1.63$ as shown later.}
$p_c \sim 0.27$, $t\sim 1.63$ as shown below.
%%SHIM_changed

As the case $\gamma_\agg = 0$, our computational results are
%%SHIM_changed
%\befor{
%that $p_c = 0.2463 \pm 0.0234$ and $\talpha = 1.769 \pm 0.154$,
%which agree with $p_c = 0.289573\pm 0.000002$ in Ref.~\ref{LZ}.
%and $\talpha=1.85 \pm 0.20$ in Refs.~\ref{FHS} and \ref{NH}.
%%and $\talpha=1.6 \pm 0.1$ in Refs.~\ref{HK} and \ref{WJC}.
%}
%\after{
$p_c = 0.273 \pm 0.012$ and $\talpha = 1.628 \pm 0.036$.
It should be noted that these computational results are obtained by FDM
method with finite lattice.
As the previous article,
we showed a finite size effect of the lattice by using an
extrapolation scheme of a lattice-size dependence of $p_c$
and $\talpha$, 
these computational results
$p_c = 0.273 \pm 0.012$ and $\talpha = 1.628 \pm 0.036$
are on the extrapolation line in Ref.~\cite{MSW};
The extrapolated values as shown
in the previous article \cite{MSW} are
$p_c^{ex} = 0.296 \pm 0.013$ and ${\talpha}^{ex} = 1.580 \pm 0.042$,
which agree with 
$p_c = 0.289573\pm 0.000002$ in Ref.~\cite{LZ}
and $\talpha=1.6 \pm 0.1$ in Refs.~\cite{HK} and \cite{WJC}.
Hence our computational scheme is consistent with these
results \cite{MSW,HK,WJC,LZ}.
%}

In this article, since we focus on the difference 
of the conductive properties among several 
$\gamma_\agg$'s, 
we basically fix the lattice-size and investigate
the conductivities in the finite region $\cB$ without
any corrections for the finite-lattice effects.
%%%SHIM_deleted

\begin{figure}[htbp]
 \begin{tabular}{cc}
  \begin{minipage}[t]{0.5\hsize}
   \begin{center}
    \includegraphics[width=\hsize, clip]{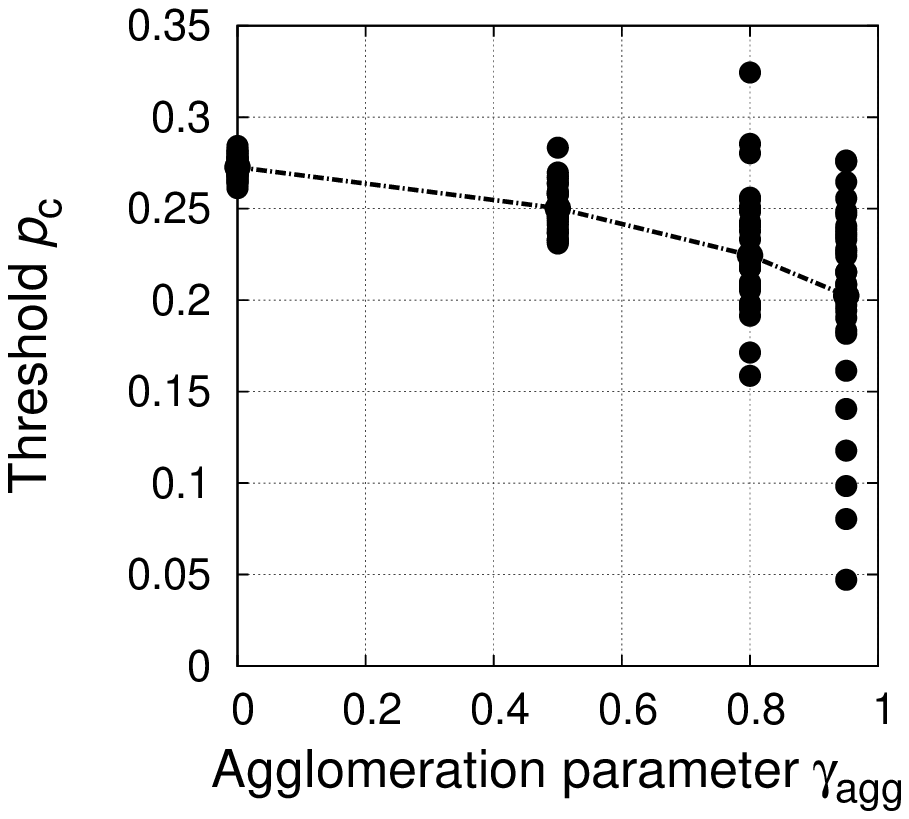}
    (a)
   \end{center}
  \end{minipage} 
  \begin{minipage}[t]{0.5\hsize}
   \begin{center}
    \includegraphics[width=\hsize, clip]{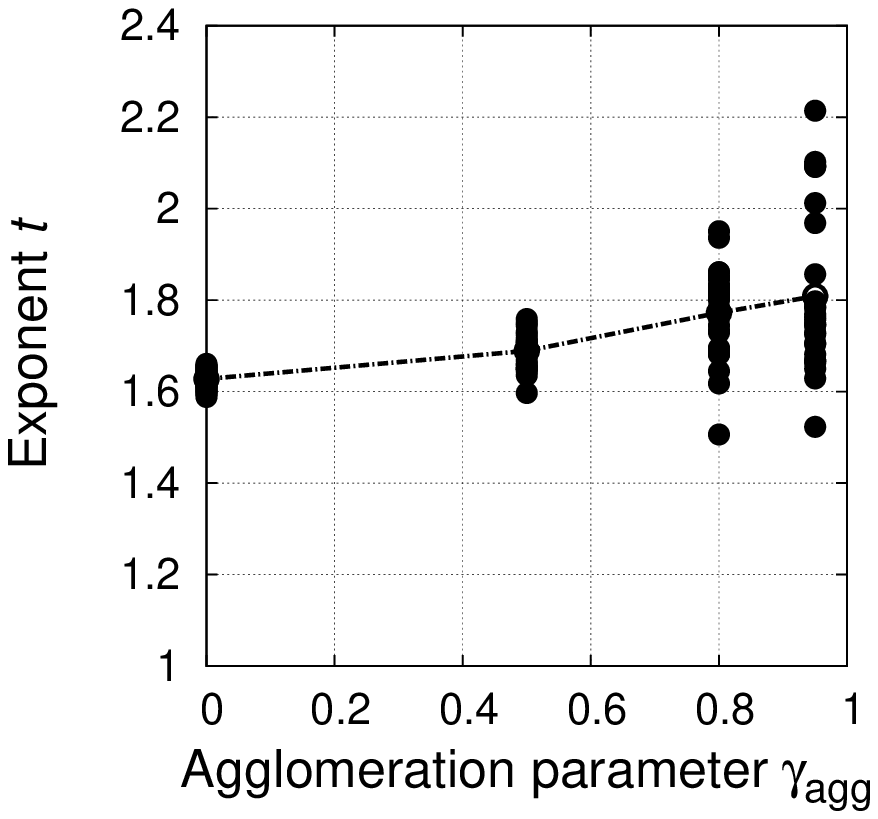}
    (b)
   \end{center}
  \end{minipage} 
 \end{tabular}
\caption{\label{fig:thandexvsag} 
The agglomeration dependence of the 
thresholds (a) and critical exponents (b). 
The filled circle corresponds to each computation for a 
%\textcolor{blue}{
seed of the pseudo-randomness
%}
and the dotted line shows the average.}
\end{figure}

The dependencies of exponents and thresholds upon the
agglomeration parameter $\gamma_\agg$ are illustrated in
Figure \ref{fig:thandexvsag} and Table \ref{tbl:thandexvsag}. 
Figure \ref{fig:thandexvsag} shows that the agglomeration
parameter $\gamma_\agg$ has large effects on the conductivity
over ACPM beyond the accuracies
%%SHIM_changed
%\befor{$\delta p_c \approx 0.01$ and $\delta \talpha \approx 0.08$.}
%\after{$\delta p_c \approx 0.005$ and $\delta \talpha \approx 0.022$.}
$\delta_\sigma p_c \approx 0.005$ and $\delta_\sigma \talpha \approx 0.022$.
%%SHIM_changed
The larger the agglomeration parameter $\gamma_\agg$ is, 
the more largely the threshold and the exponent depend on the seed $i_s$.
It means that the 
%\textcolor{red}{\sout{statistical error}} 
%\textcolor{blue}{
variance
%} 
%\textcolor{red}{\sout{coming from the randomness}}
is enhanced
by the agglomeration parameter $\gamma_\agg$.

Furthermore the larger the agglomeration parameter $\gamma_\agg$ is,
the smaller the trend of the threshold is and the larger 
that of the exponent is.
As shown in Table \ref{tbl:thandexvsag}, 
the larger the agglomeration parameter $\gamma_\agg$ becomes,
the smaller the average of the thresholds $\overline{p_c}(\gamma_\agg)$ is 
and the larger the average of the exponent $\overline{t}(\gamma_\agg)$ is.

\begin{table}[htbp]
\begin{center}
\caption{The $\gamma_\agg$ dependence of the threshold and the exponent.}
\label{tbl:thandexvsag}
\begin{tabular}{|c|c|c|c|c|c|c|c|c|c|c|}
\hline
\multicolumn{1}{|c|}{ } & 
\multicolumn{3}{|c|}{Threshold}& 
\multicolumn{3}{|c|}{Exponent} \\
\hline
$\gamma_\agg$ &Average & Maximum & Minimum & Average & Maximum & Minimum  \\
\hline
0.0  & 0.273 & 0.284 & 0.261  & 1.628 & 1.661 & 1.588 \\
0.5  & 0.250 & 0.283 & 0.231  & 1.689 & 1.758 & 1.597 \\
0.8  & 0.224 & 0.324 & 0.159  & 1.772 & 1.951 & 1.506 \\
0.95 & 0.202 & 0.276 & 0.047  & 1.809 & 2.214 & 1.523 \\
\hline
\end{tabular}
\end{center}
\end{table}

%\textcolor{blue}{
Figures \ref{fig:pcurve_allseeds_1} and \ref{fig:pcurve_allseeds_2}
show that 
the dependence of agglomeration parameters $\gamma_\agg$
on the conductivity curves are enhanced under the threshold $p_c$
for the logarithm scale.
Under the percolation threshold $p_c$, our computation can read the
computation of dielectric behavior on a
random configuration of metal particles in
the dielectric matter, which was reported in Refs.~\cite{GB}
and \cite{SMKK} if we consider the conductivity as
the dielectric constant. 
Physically speaking,
the behavior is related to the electric breakdown.
Though the differences $p < p_c$ 
have 
%\textcolor{red}{\sout{very small}}
%\textcolor{blue}{
negligible
%}
 effects on the estimations of the 
conductivity curves (\ref{eq:sigma_xi}) due to
the fitting method for the linear scaling,
they might be crucial for the dielectric behavior.
%}

\section{Discussion on the agglomeration effects}

We now consider the reason why the agglomeration has the effects on the
conductivity in our ACPM.
%\befor{
%Since we have dealt with the percolation phenomena on a finite region $\cB$,
%and the agglomeration must have an influence on 
%the size of the percolation cluster, we first review the size effect on
%the conductivity following Ref.~\ref{AS} and simply consider
%it in CPM numerically.}

%\textcolor{blue}{
We have dealt with the percolation phenomena on a finite region $\cB$
whereas it is well-known that the conductivity
has dependence on the size of region as  in Chapter 5.1 in Ref.~\cite{SA}.
We naturally have  the
characteristics length $r_c$ induced from 
the size of the particles or the radius of the particle $\rho$;
for vanishing $\gamma_\agg$, 
i.e., $r_c(\gamma_\agg=0) = \rho$. 
Due to the finiteness of size $L$ of $\cB$, 
the ratio $L/r_c$ has an effect on
the conductivity;
at $\gamma_\agg=0$, we have that $L/r_c \approx 36$.
For non-vanishing agglomeration parameter $\gamma_\agg\neq0$,
it is expected that the characteristic length $r_c$ is larger than that of 
CPMs ($\gamma_\agg=0$) due to the agglomeration, 
i.e., $r_c(\gamma_\agg\neq0) > \rho$.
Further by letting  $\xi=\xi(p,\gamma_\agg)$ be the correlation length
which represents the percolation phenomenon,
we should compare the size of region $L$, the correlation
length $\xi$ and the characteristic length $r_c$. 
According to Chapter 5.1 in Ref.~\cite{SA},
the total conductivity behaves like
\begin{equation}
	\sigma_\total(p, \gamma_\agg) \propto 
    \left\{ \begin{array}{ll}
         (\xi/r_c)^{-\mu}, & \mbox{for} \ L/r_c \gg (\xi/r_c),\\
         (L/r_c)^{-\mu}, & \mbox{for} \ L/r_c \ll (\xi/r_c),\\
      \end{array} \right.
\label{eq:sigma_xi}
\end{equation}
where $\mu$ is the non-negative parameter (73a) in Ref.~\cite{SA}.
Since 
it is known that $\xi$ diverges at the critical point $p_c$
and in every numerical estimation we handle only a finite region,
the formula (\ref{eq:sigma_xi}) means that
every numerical estimation gives higher conductivity in the vicinity 
of the point $p_c$ than that of an infinite region apart from the
%\textcolor{blue}{
variances.
%}
%\textcolor{blue}{ \sout{random fluctuation}.  }

%\textcolor{blue}{
Though the percolation theory is basically
concerned with conductivity curve in the infinite region,
we concern ourselves about the size effect of the real materials.
Further in Ref.~\cite{MSW}, we have showed that the shape effect 
has crucial effects on the conductivity curves.
%fractional value model of CPMs as we will
%show it in detail in later.
Thus we will consider the geometrical properties of agglomerated clusters.
%}

\subsection{The geometrical features in ACPM}\label{subsec:4.3}

Let us consider the size and the shape 
of the agglomerated clusters which depend
on the agglomeration parameter $\gamma_\agg$.

\subsubsection{The characteristic length $r_c$ in ACPM at $p=0.1$}
\label{subsubsec:4.1.1}

%\del{
%We return to consider one of the effects of the agglomeration on the
%conductivity in our ACPM with $\rho=1$
%as the size effect and give an estimation on 
%the characteristic length $r_c$ for non-vanishing $\gamma_\agg$.
%}
%\ins{
First we consider the size effect and give an estimation on 
the characteristic length $r_c$ for non-vanishing $\gamma_\agg$.
Since
%}
it is difficult to estimate it,
we compute the difference among the size of the isolated
agglomerated clusters at a lower volume fraction $p$ than
$p_c$. In other words, we compute 
statistical behavior of the percolation clusters,
or the connected particles, at $p=0.1$.

When the agglomeration parameter $\gamma_\agg$ becomes large, 
it is expected that the size
of the percolation cluster becomes larger than one of the uniform
randomness or the case ($\gamma_\agg = 0$).
The size of the percolation cluster is associated with the
correlation length $\xi(p, \gamma_\agg)$ in (\ref{eq:sigma_xi}).

\begin{figure}[htbp]
 \begin{center}
  \includegraphics[scale=0.9, clip]{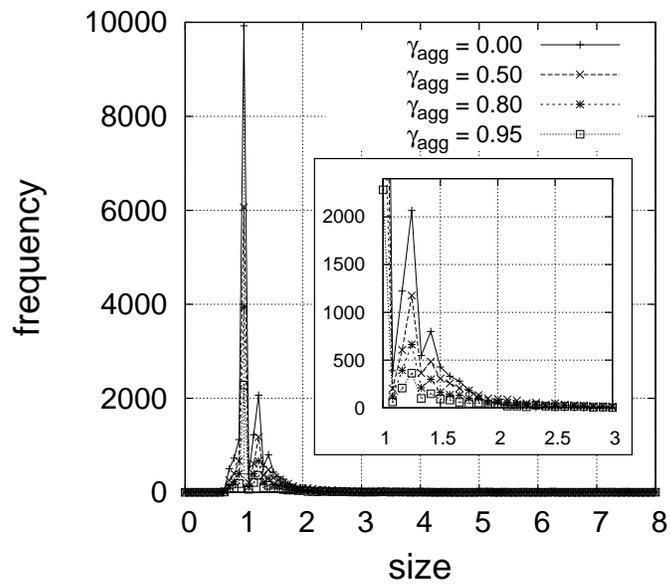}
 \end{center}
\caption{\label{fig:clustersize}
The histogram of size $\rho_\clst$ of percolation cluster for
each agglomeration parameter  $\gamma_\agg$ at $p=0.1$.}
\end{figure}

In order to evaluate the effect of the size,
%%SHIM_ADDED
%\ins{
we consider effective radius $\rho_\clst$ and
maximum length $L_\max$ of each agglomerated clusters in ACPM, 
and the numbers $N_\agg$ of agglomerated clusters. 
Here the effective radius $\rho_\clst$ is defined such that
$4\pi \rho_\clst^3/3$ is equal to the  volume of the percolation cluster.
First, we consider a
histogram of 
the effective radius $\rho_\clst$.
Figure \ref{fig:clustersize} shows frequency of 
the effective radius $\rho_\clst$ at
the volume fraction $p=0.1$ which is smaller than any thresholds 
$\overline{p_c}(\gamma_\agg)$.
Figure \ref{fig:clustersize} means that the larger 
%\textcolor{blue}{
the agglomeration parameter $\gamma_\agg$
%}
is, the larger the size of the cluster is.

Figure \ref{fig:clustersize} shows the fact that the histogram has
two peaks; the first peak at $\rho_\clst=\rho$ as the size of
isolated particle, and the second peak around 
$\rho_\clst=\sqrt[3]{2}\approx 1.26\rho$ 
as the size of two particles. 
The probability of two slightly connected
particles is larger than the probability of the state
which has $\rho_\clst \in (\rho, \sqrt[3]{2}\rho)$ 
because of radial measure $\rho_\clst^2 d\rho_\clst$ for 
the radius $\rho_\clst$ from the center of a particle.
The percolation is determined by the largest
cluster size but it should be statistically treated.
We consider the right hand side of the second peak of the histograms.
We fit the shape of the histogram over $\rho_\clst\in [1.3, 8]$ by 
$h(\rho_\clst) = A\exp(-(\rho_\clst-\rho)/\rho_\clst^0)$ well 
using the least mean square error method,
where $A$ and $\rho_\clst^0$ are fitting parameters.
Then $\rho_\clst^0$ is given in Table \ref{tbl:rhovsag}.
Here $\delta h$ is the square root of the average of the 
least mean square error.

\begin{table}[htbp]
\begin{center}
\caption{The $\rho_\clst$ dependence on 
the agglomeration parameter $\gamma_\agg$.}
\label{tbl:rhovsag}
\begin{tabular}{|c|r|r|r|r|}
\hline
$\gamma_\agg$       & 0     & 0.5   & 0.8   & 0.95  \\
\hline
$\rho_\clst^0/\rho$ & 0.315 & 0.458 & 0.476 & 0.522 \\
\hline
$\delta h$          & 0.020 & 0.019 & 0.024 & 0.028 \\
\hline
\end{tabular}
\end{center}
\end{table}

Table \ref{tbl:rhovsag} shows that the agglomeration parameter
$\gamma_\agg$ increases the size of the clusters at $p=0.1$.
The characteristic length $r_c$ is directly relevant to 
the size of clusters $\rho_\clst^0$.
Though it is difficult to evaluate the difference of the size of clusters
$\rho_\clst^0$ and also $r_c$
at $p \ge p_c$, it is expected that it plays similar roles even for
every $p \in [0,1]$. 

\begin{table}[htbp]
\begin{center}
\caption{The $\overline{\rho_\clst}$, $\overline{L_\max}$, and $N_\agg$
of the agglomerated ($\rho_\clst > 1.3$) particles dependence on
the agglomeration parameter $\gamma_\agg$ at $p=0.1$.}
\label{tbl:averhovsag}
\begin{tabular}{|c|r|r|r|r|}
\hline
$\gamma_\agg$           & 0     & 0.5   & 0.8   & 0.95  \\
\hline
$\overline{\rho_\clst}$ & 1.572 & 1.800 & 1.978 & 2.200 \\
\hline
$\overline{L_\max}/2$   & 2.803 & 3.310 & 3.668 & 4.009 \\
\hline
$N_\agg$                & 2977  & 2725  & 1722  & 952   \\
\hline
\end{tabular}
\end{center}
\end{table}

%\textcolor{blue}{
Further Table \ref{tbl:averhovsag} shows the averages of the size
$\overline{\rho_\clst}$ and the maximum distance $\overline{L_\max}$ of
the percolation clusters which have $\rho_\clst > 1.3 \rho$,
and their number $N_\agg$. 
The larger $\gamma_\agg$ is,
the larger $\overline{L_\max}$ and $\overline{\rho_\clst}$ are,
and the smaller $N_\agg$ is.
%}

They mean that the larger $\gamma_\agg$ is, the smaller 
the ratio $L/r_c$ is.
Hence it is expected that
 the larger $\gamma_\agg$ is, the larger 
%\textcolor{blue}{
the variance becomes.
%}
%\textcolor{red}{\sout{the fluctuation becomes}}.

\subsubsection{The shape of the agglomerated cluster in ACPM}
\label{subsubsec:4.1.2}

Here we consider the shape of the agglomerated cluster.
In our algorithm which is shown in Figure \ref{fig:flowchart},
a new particle with $\gamma < \gamma_\agg$ must be connected with
the particles which have been already placed. 
If the volume fraction is much less
than the threshold, the connected (agglomerated) cluster which 
consists of $N$ particles
could be regarded as an orbit of a random walk for discrete
$N$ time step.

\begin{figure}[htbp]
 \begin{tabular}{cc}
  \begin{minipage}[t]{0.5\hsize}
   \begin{center}
    \includegraphics[width=\hsize, clip]{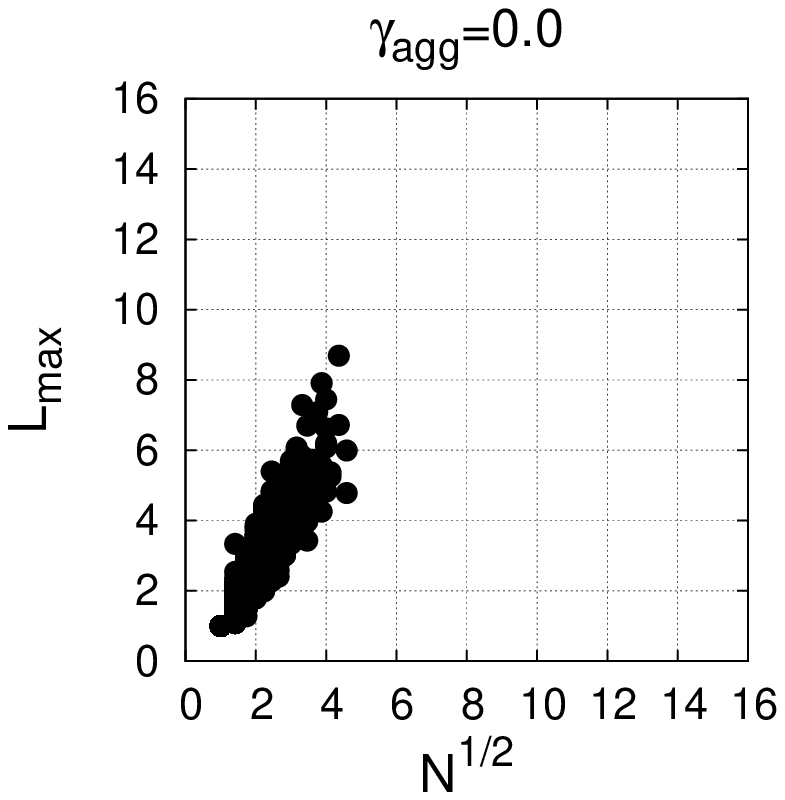}
    (a)
   \end{center}
  \end{minipage} 
  \begin{minipage}[t]{0.5\hsize}
   \begin{center}
    \includegraphics[width=\hsize, clip]{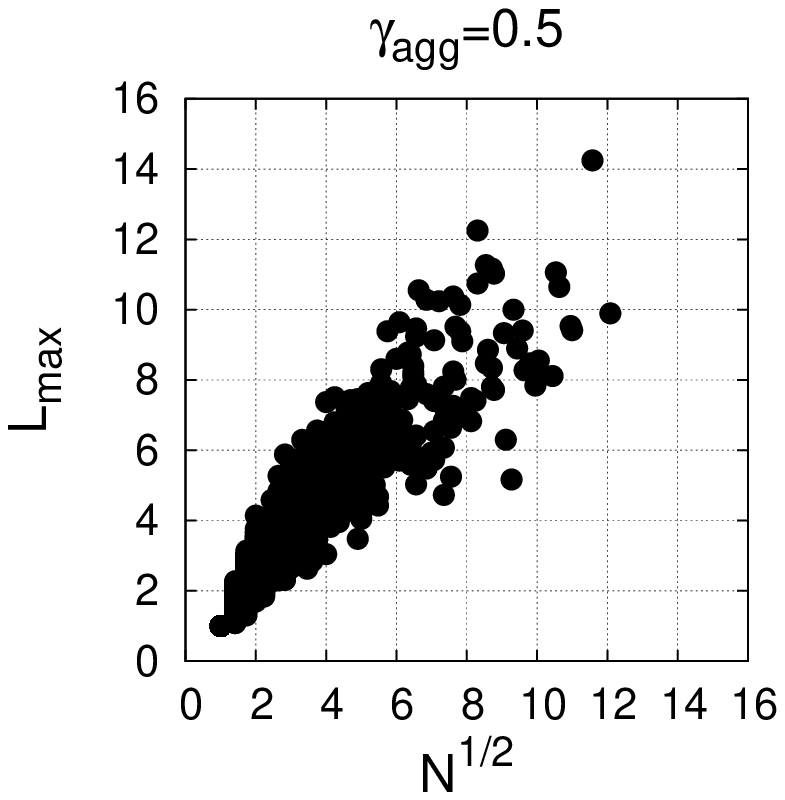}
    (b)
   \end{center}
  \end{minipage} 
 \end{tabular}
\caption{\label{fig:sqrtNvsLmax} 
The dependence of the $L_\max$ on the $\sqrt{N}$ for
 $\gamma_\agg = 0.0, 0.5$.
%}
}
\end{figure}

In fact, the size of the agglomerated cluster is proportional to
the square root of $N$ as shown in Figure \ref{fig:sqrtNvsLmax}.
Figure \ref{fig:sqrtNvsLmax} displays the correlation between the $L_\max$
and $\sqrt{N}$ for $\gamma_\agg = 0.0, 0.5$. They are linearly
relative.
This coincides with the properties of the random walk
due to the central limit theorem \cite{Fel}.
%}
%%MATSU_added_SHIM_modified

Figure \ref{fig:sqrtNvsLmax} means that if we regard the agglomerated
cluster as a cylinder, the radius around the center axis is proportional
to $\sqrt[4]{N}$ since the volume should be proportional
to $N$ but $L_\max \propto \sqrt{N}$.
For a sufficiently large $N$, the agglomerated cluster might be regarded
as a thin cylinder rather than a thick cylinder. 
%The behavior can be regarded as
%broad distribution continuum percolation model of Hara-Odagiri
%and thus it is expected that the thinner shape is, 
%the larger exponents and the smaller thresholds also have. 
%}

%%SHIM_added

\subsection{The agglomeration effect on conductivity in ACPM}
\label{subsec:4.1}

In this section, we will investigate the behavior in
Figure \ref{fig:thandexvsag} as 
the agglomeration effect on conductivity in ACPM.

As shown above,
when we fix $\cB$, the  larger $r_c(\gamma_\agg)$ becomes,
the smaller the effective size $L/r_c$ in (\ref{eq:sigma_xi})
is regarded.
It is expected that
the total conductivity $\sigma_\total(p, \gamma_\agg, i_s)$
strongly depends on the configuration $\Rconf_{\gamma_\agg, p,i_s}$
since statistical average does depend upon the ratio $L/r_c$
which is not sufficiently large for $\gamma_\agg>0$.
As shown in Figure \ref{fig:thandexvsag},
it means that the larger $\gamma_\agg$ is,
the more largely the threshold
$p_c(\gamma_\agg\neq0, i_s)$ and the exponent
$\talpha(\gamma_\agg\neq0, i_s)$ depend on the seed $i_s$.
Thus we consider the size effect first.

%Figure 8 exhibits the correlation between the thresholds and 
%the exponents in both the agglomeration effects.
%%The smaller $\overline{p_c}$ has, the larger exponent $\talpha$ is.
%The smaller $\overline{p_c}$ has, the larger 
%statistical average of the exponent $\overline{\talpha}$ is.

\subsubsection{The size effect on conductivity in CPM}
\label{subsec:4.2}

In this subsection
we consider the size effect on conductivity in CPM 
or the case $\gamma_\agg=0$ by changing the radius $\rho$
directly.

We computed the threshold $p_c$ and the exponent $\talpha$ for the
different radius $\rho(>1)$  by fixing the size of $\cB$.
(From the numerical viewpoint, we performed the similar computations
of $\gamma_\agg=0$ for small $N_x=N_y=N_z$ by fixing the mesh of
$\rho$. It corresponds to the variation of the radius $\rho$ of
the particles for fixing $\cB$ relatively.)

\begin{figure}[htbp]
 \begin{tabular}{cc}
  \begin{minipage}[t]{0.5\hsize}
   \begin{center}
    \includegraphics[width=\hsize, clip]{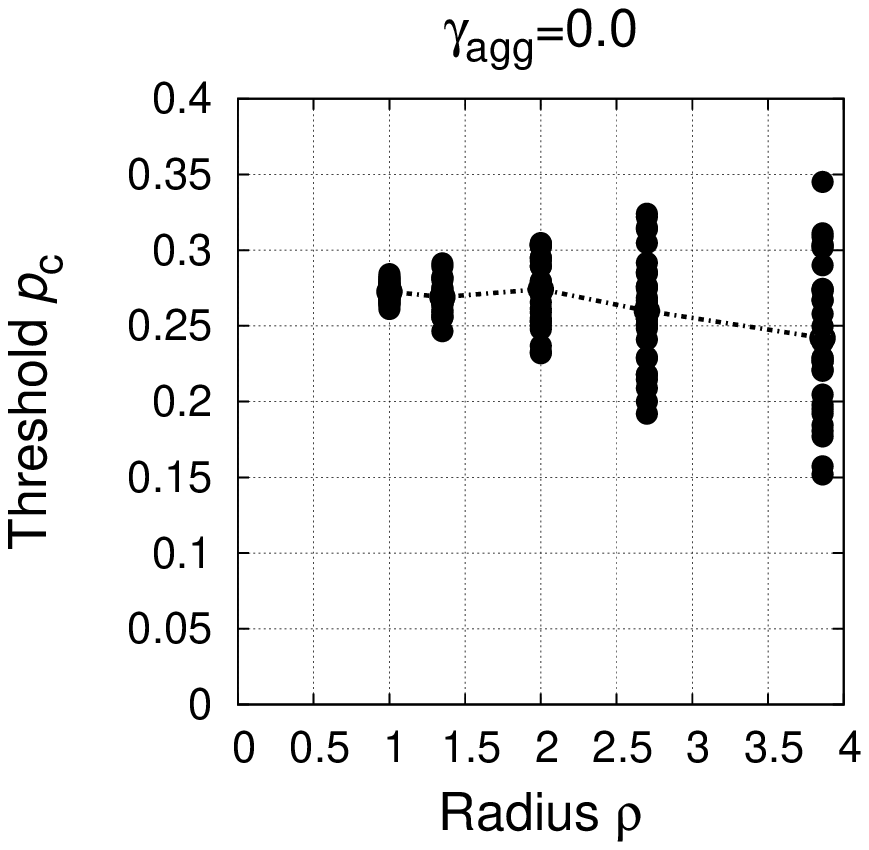}
    (a)
   \end{center}
  \end{minipage} 

  \begin{minipage}[t]{0.5\hsize}
   \begin{center}
    \includegraphics[width=\hsize, clip]{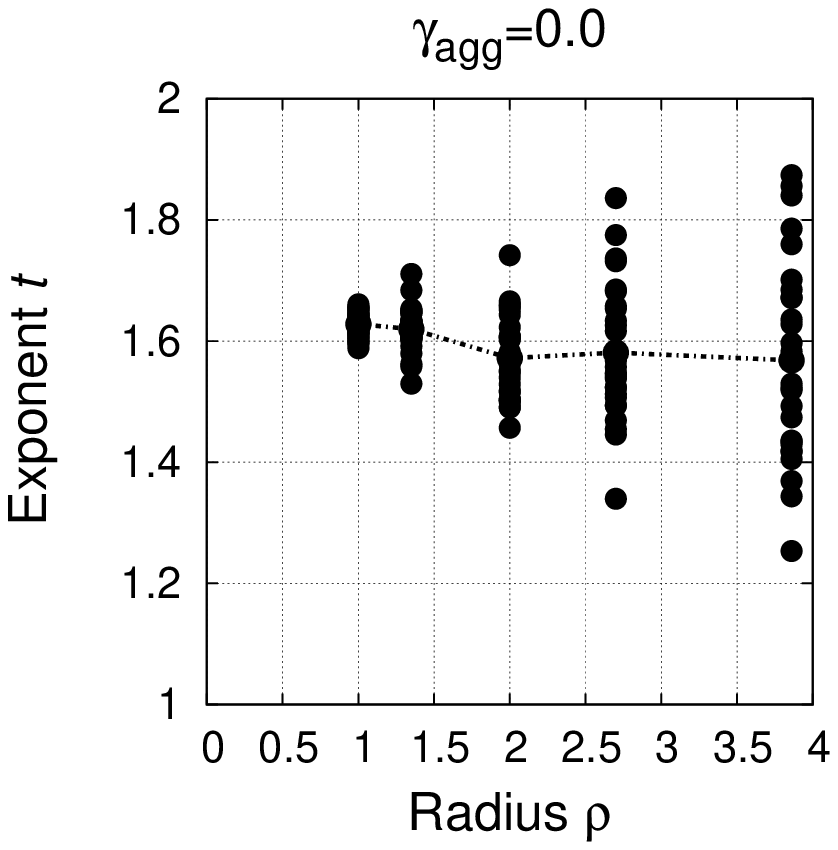}
    (b)
   \end{center}
  \end{minipage} 
 \end{tabular}
\caption{\label{fig:largeparticle} 
Thresholds (a) and critical exponents (b)
vs radius of particles in the same $\cB$.
The filled circle corresponds to each computation for a seed
and the dotted line is their average.}
\end{figure}

\begin{figure}[htbp]
 \begin{tabular}{ccc}
  \begin{minipage}[t]{0.33\hsize}
   \begin{center}
    \includegraphics[width=\hsize, clip]{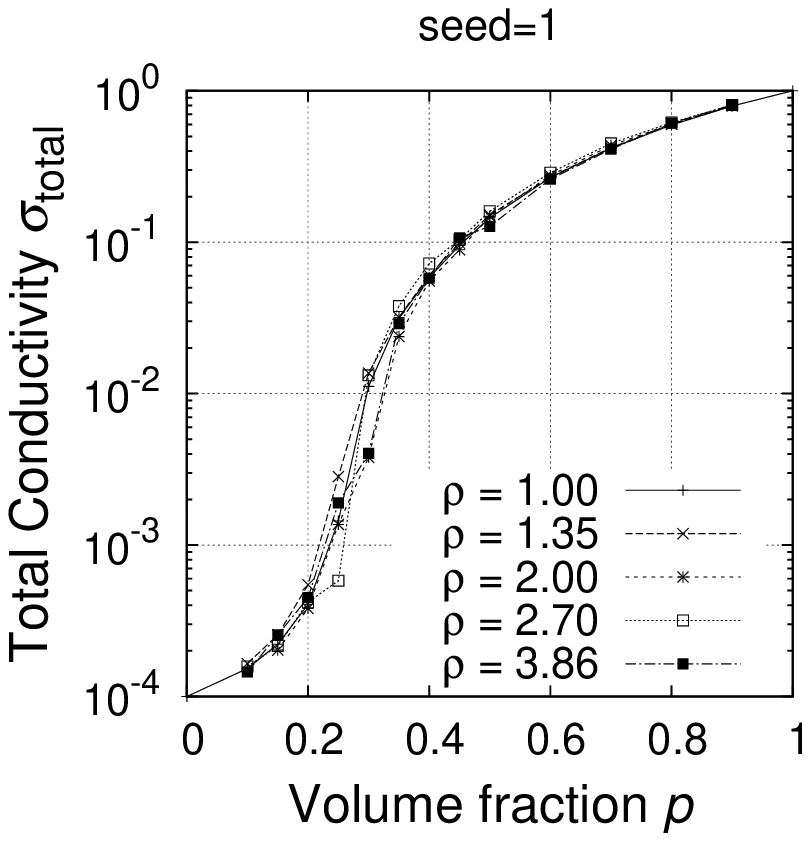}
    \includegraphics[width=\hsize, clip]{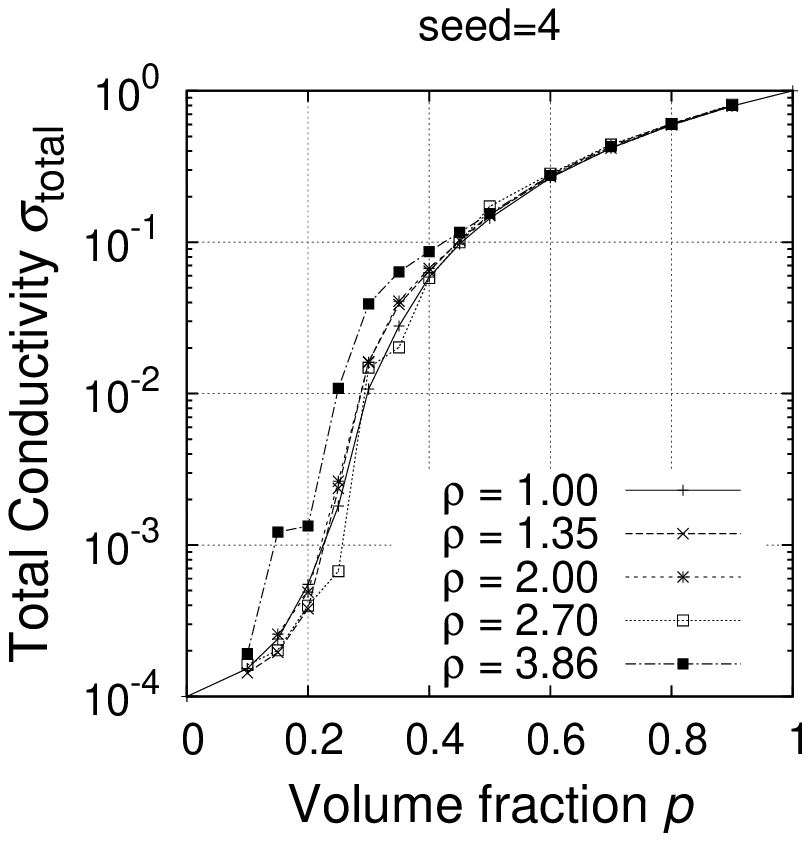}
    \includegraphics[width=\hsize, clip]{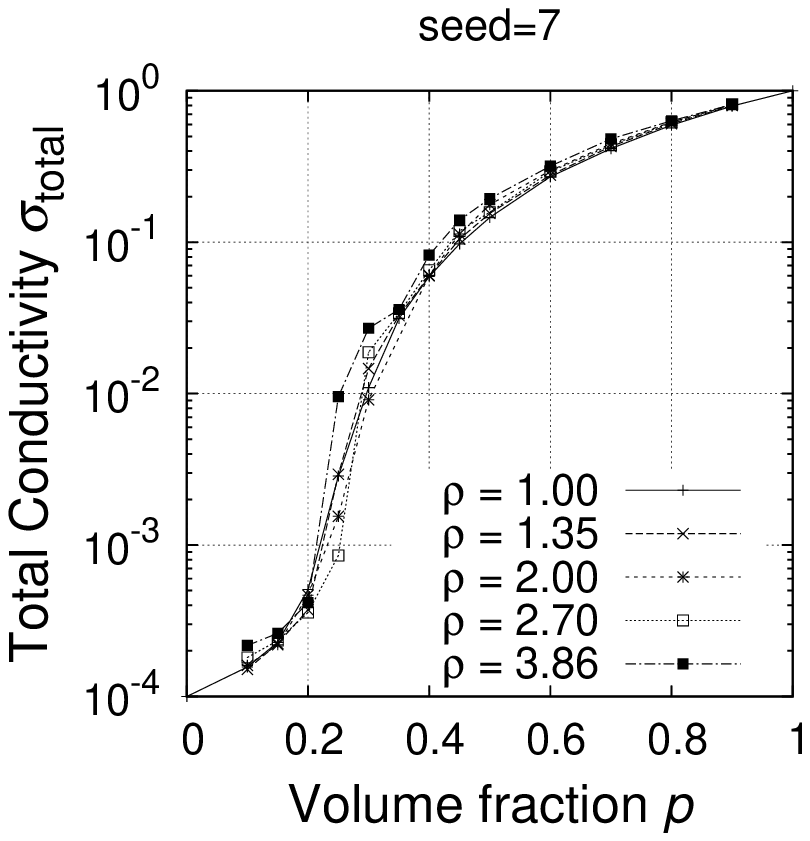}
   \end{center}
  \end{minipage} 
  \begin{minipage}[t]{0.33\hsize}
   \begin{center}
    \includegraphics[width=\hsize, clip]{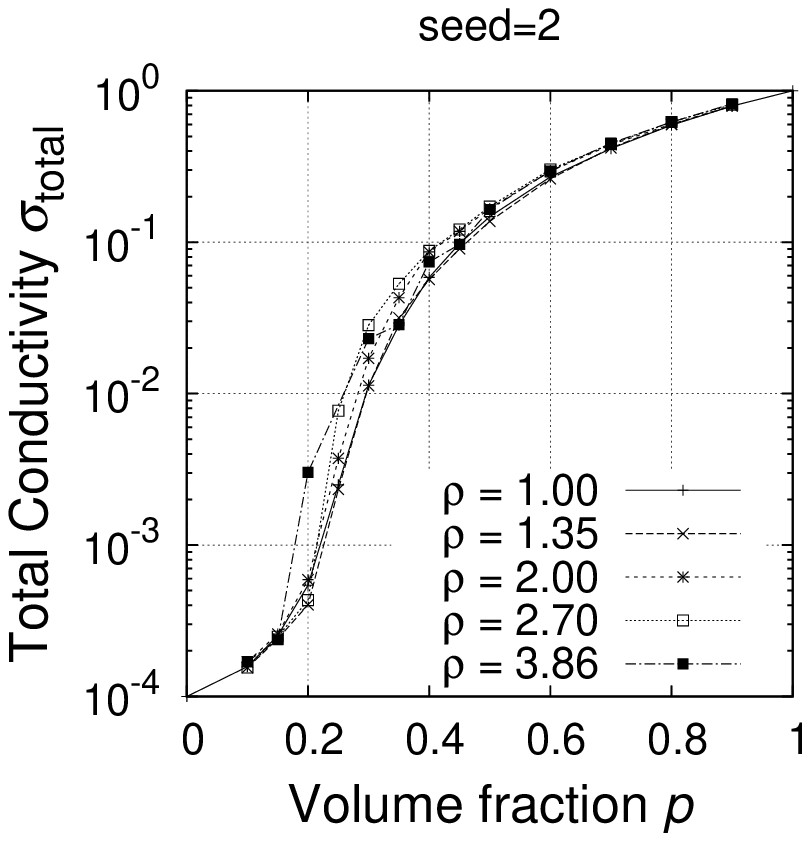}
    \includegraphics[width=\hsize, clip]{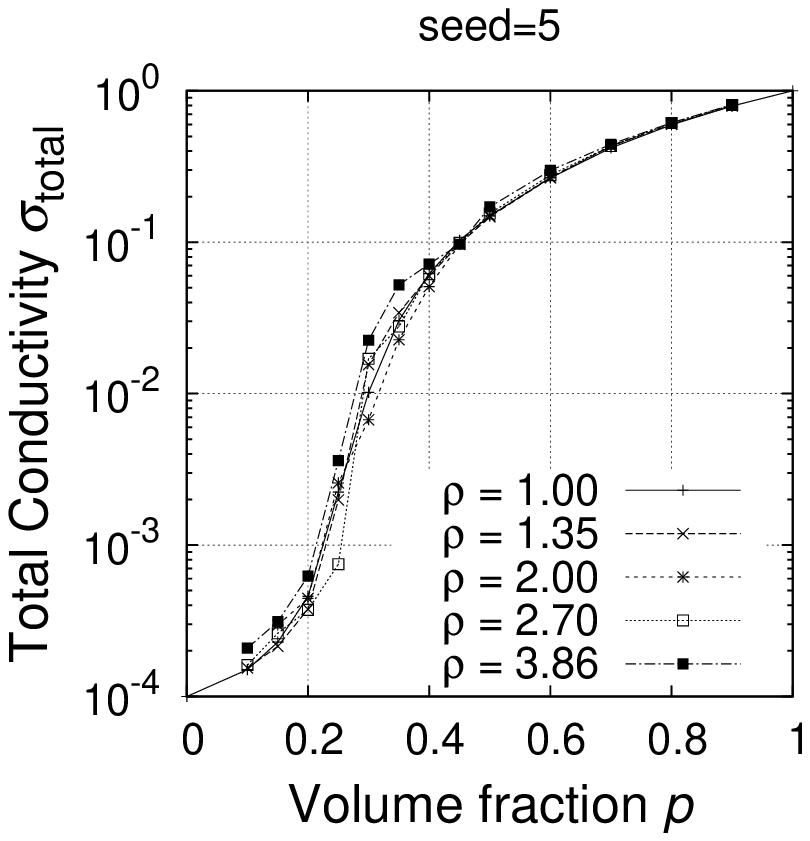}
    \includegraphics[width=\hsize, clip]{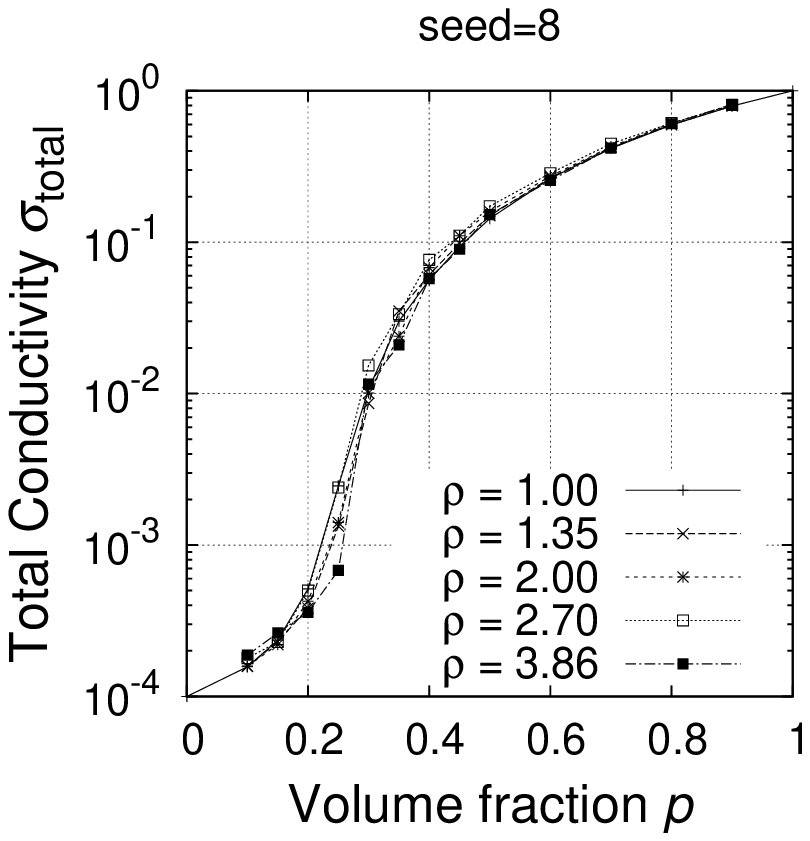}
   \end{center}
  \end{minipage} 
  \begin{minipage}[t]{0.33\hsize}
   \begin{center}
    \includegraphics[width=\hsize, clip]{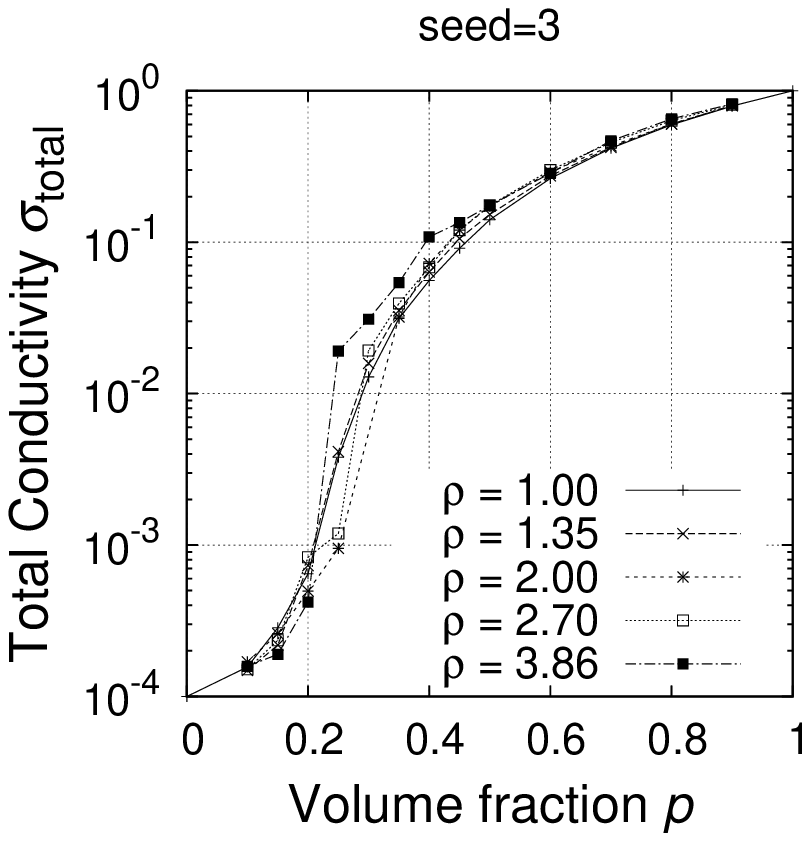}
    \includegraphics[width=\hsize, clip]{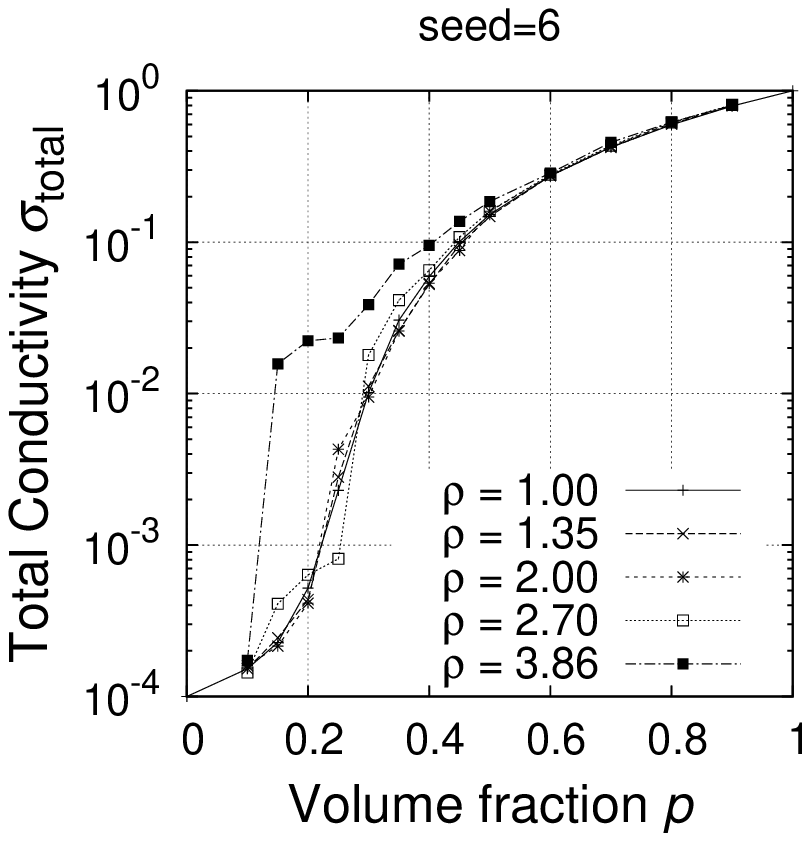}
    \includegraphics[width=\hsize, clip]{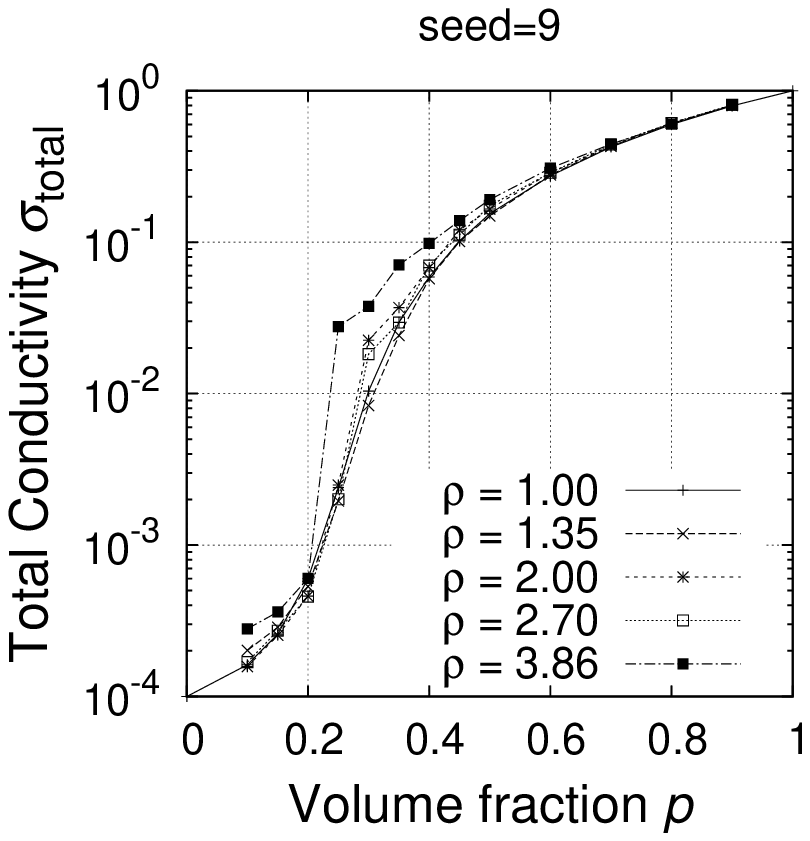}
   \end{center}
  \end{minipage} 
 \end{tabular}
\caption{\label{fig:largeparticle2} The conductivity curves for
the radius $\rho = 1.0, 1.35, 2.0, 2.7,$ and $3.86$ in the same $\cB$
for the seed $i_s=1, 2, \dots, 9$ 
%\textcolor{blue}{
of the pseudo-randomness.
%} 
}
\end{figure}

%%%%%%%%%%%%%%%%%%%%%%%%%%%%%%%%%%%%%%%%%%%%%%%%%%%%%%
Figure \ref{fig:largeparticle2} illustrates
the dependence of the particle radius $\rho$ on the conductivity
curves for the logarithm scale for nine cases, though
we computed thirty cases for each $\rho$.
%%%%%%%%%%%%%%%%%%%%%%%%%%%%%%%%%%%%%%%%%%%%%%%%%%%%%%
Figure \ref{fig:largeparticle} and Table \ref{tbl:largeparticle}
give the dependence of the threshold and the exponent
on the size of particles for the same $\cB$.
Figure \ref{fig:largeparticle} shows that
the (relatively) larger the radius $\rho$ is,
the larger 
%\textcolor{red}{\sout{the fluctuations}}
%\textcolor{blue}{
the dispersions
%}
of the threshold
$p_c$ and the exponent $\talpha$ are.
This property is similar to Figure \ref{fig:thandexvsag}.
However
%\textcolor{red}{\sout{the fluctuations}}
%\textcolor{blue}{
the dispersions
%}
of both threshold and exponent in
Figure \ref{fig:largeparticle}
look symmetric with respect to their averages
whereas 
%\textcolor{red}{\sout{the fluctuations}}
%\textcolor{blue}{
the dispersions
%}
in Figure \ref{fig:thandexvsag}
show asymmetry.
%}

\begin{figure}[htbp]
 \begin{tabular}{cc}
  \begin{minipage}[t]{0.5\hsize}
   \begin{center}
    \includegraphics[width=\hsize, clip]{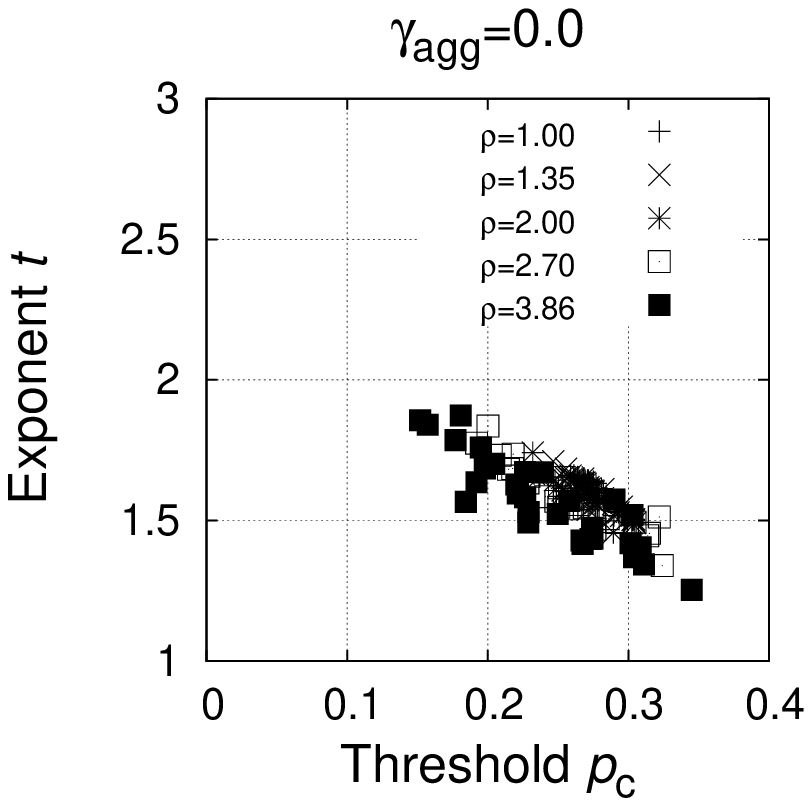}
    (a)
   \end{center}
  \end{minipage} 
  \begin{minipage}[t]{0.5\hsize}
   \begin{center}
    \includegraphics[width=\hsize, clip]{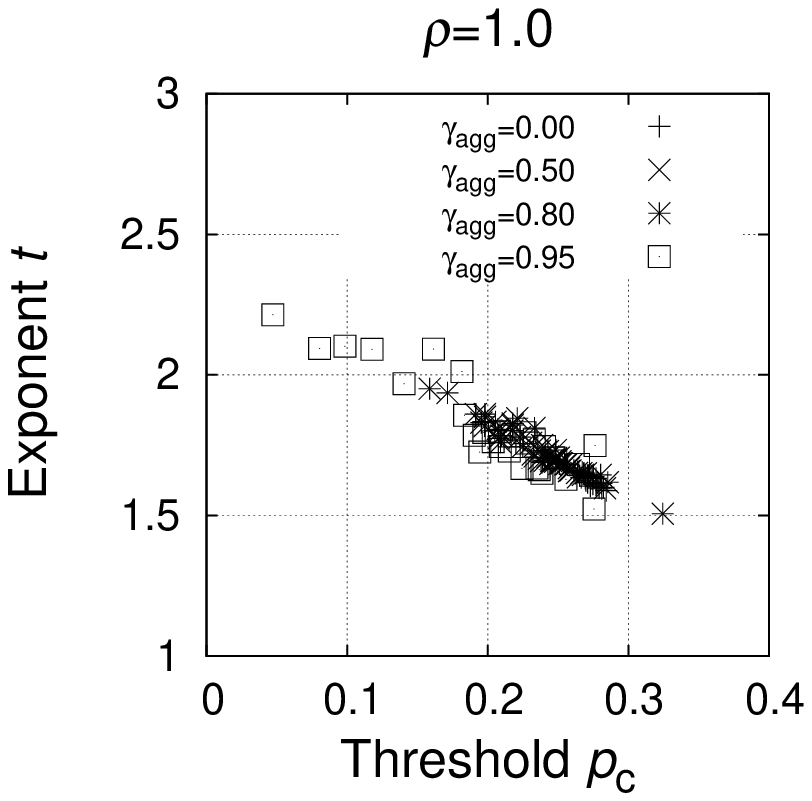}
    (b)
   \end{center}
  \end{minipage} 
 \end{tabular}
\caption{\label{fig:expvsthr} 
Critical exponents vs thresholds 
(a) for various $\rho$ with $\gamma_\agg=0.0$,
and (b) for various $\gamma_\agg$ with $\rho=1$.
}
\end{figure}

The equation (\ref{eq:sigma_xi}) is evaluated from the
statistical viewpoint. 
There is no difference among $\rho$'s if $\cB$ has infinite region.
Since $\cB$ has a finite size, 
for large $\rho$,
we don't have so sufficiently many
particles in $\cB$ that its statistical average works well.
Hence 
%\textcolor{red}{\sout{the fluctuation}}
%\textcolor{blue}{
the deviation
%}
is enhanced.
In other words, the dependence of the conductivity curve 
(\ref{eq:sigmatotal}) on the seed $i_s$ 
%\textcolor{blue}{
of the pseudo-randomness
%} 
is larger than
that of $\rho = 1$ as in Figure \ref{fig:largeparticle}.
%}

Since (\ref{eq:sigma_xi}) means that the finiteness of
$\cB$ makes the threshold $p_c$ small statistically,
the trend of the statistical average of the threshold shows 
that  the larger the size $\rho$ in CPM is, 
the smaller the threshold $\overline{p_c}(\rho)$ is  as shown in
Table \ref{tbl:largeparticle} and Figure \ref{fig:largeparticle} (a).
Here, $\overline{p_c}(\rho)$ stands for the statistical
average of the $p_c(\rho, i_s)$ over 
the seed $i_s$ 
%\textcolor{blue}{
of the pseudo-randomness.
%} 
This property of the threshold is also similar to 
Figure \ref{fig:thandexvsag} (a),
though the dependence of the exponents in Figure \ref{fig:largeparticle}
(b) is quite different from Figure \ref{fig:thandexvsag} (b).

When the individual
total conductivity $\sigma_\total(p,\rho,i_s)$ has smaller 
$p_c(\rho, i_s)$ than
$\overline{p_c}(\rho)$,
the non-vanishing $\sigma_\total(p, \rho,i_s)$ at $p \in
(p_c(\rho, i_s), \overline{p_c}(\rho))$,
must increase weakly with respect to $p$.
It means that the exponent $\talpha(\rho, i_s)$ 
becomes larger in the conductivity
curve (\ref{eq:sigmatotal}) than 
%\textcolor{blue}{
its average
%}
$\overline{\talpha}$.
It implies that
the exponent $\talpha(\rho,i_s)$ and the threshold $p_c(\rho,i_s)$ 
in CPMs of different $\rho$ and $i_s$ are correlative and
that the smaller the thresholds $p_c(\rho,i_s)$ are, the
larger the exponents $\talpha(\rho,i_s)$ are. 
Particularly
Figure \ref{fig:expvsthr}(a) exhibits the correlation between the exponent
$t$ and the threshold $p_c$ in CPMs.
%}

%Hence the tendency obeying (\ref{eq:sigma_xi})
%are also inherent in Figure \ref{fig:largeparticle}, Figure 10 (b)
%and Table 5.

\begin{table}[htbp]
\begin{center}
\caption{The $\rho$ dependence 
of the threshold and the exponent.}
\label{tbl:largeparticle}
\begin{tabular}{|c|c|c|c|c|c|c|c|c|c|c|}
\hline
\multicolumn{1}{|c|}{ } & 
\multicolumn{3}{|c|}{Threshold}& 
\multicolumn{3}{|c|}{Exponent} \\
\hline
$\rho$ &Average & Maximum & Minimum & Average & Maximum & Minimum  \\
\hline
1.0    & 0.273  & 0.284   & 0.261   & 1.628   & 1.661   & 1.588 \\
1.35   & 0.269  & 0.291   & 0.246   & 1.620   & 1.711   & 1.530 \\
2.0    & 0.274  & 0.305   & 0.232   & 1.572   & 1.742   & 1.457 \\
2.7    & 0.260  & 0.324   & 0.192   & 1.581   & 1.836   & 1.340 \\
3.86   & 0.242  & 0.345   & 0.152   & 1.568   & 1.873   & 1.253 \\
\hline
\end{tabular}
\end{center}
\end{table}

\subsubsection{The size effect on conductivity in ACPM}
\label{subsec:4.4}

Following the above discussions, we consider the agglomeration
effect on the conductivity in our ACPM.

Subsection \ref{subsec:4.3}
 means that the larger the agglomeration parameter $\gamma_\agg$ is,
the larger the size of the percolation clusters and characteristic length
$r_c$ are and the smaller the effective size $L/r_c$ in
(\ref{eq:sigma_xi}) is.
For the agglomeration parameter $\gamma_\agg$ deviated from $0$, 
$\cB$ corresponds to a relatively smaller region than
the uniform random case ($\gamma_\agg = 0$). 
Subsection \ref{subsec:4.2}
means that 
due to the finite size effect,
the dependence of the threshold and the exponent upon the seed of the
pseudo-randomness, i.e., 
%\textcolor{red}{\sout{these fluctuations}}
%\textcolor{blue}{
these variances
%}
are larger than the case of $\gamma_\agg = 0$. 
In other words, the large 
%\textcolor{red}{\sout{fluctuation}}
%\textcolor{blue}{
deviation
%}
for non-vanishing $\gamma_\agg$
comes from the finiteness of $\cB$ and the size of the agglomerated
clusters.

%the threshold becomes
%smaller in our evaluation. 
%It is expected that
%the trend of the threshold $p_c$ to $\gamma_\agg > 0$ 
%is smaller than $p_c$ at the
%case $\gamma_\agg = 0$
% even though there is a large fluctuation of $p_c$.

%\textcolor{blue}{
The dependence of the particle radius $\rho$ on the conductivity
curves for the logarithm scale is displayed in Figure \ref{fig:largeparticle2}.
The variance of the curves look enhanced under the threshold $p_c$.
The same behavior is observed in Figures \ref{fig:pcurve_allseeds_1}
and \ref{fig:pcurve_allseeds_2}; the dependence of the agglomeration
parameter $\gamma_\agg$ on the conductivity curves.
It means that
they also show the relation between the size effect 
and the agglomeration effect, as we mentioned above.
Since the smaller $p$ is, the smaller the number of the clusters is,
these effects becomes evident for small $p$.
%}

Since the agglomeration makes the size of the characteristic
length $r_c$ larger than that of non-agglomeration state,
it is expected that 
%\ins{
$\overline{p_c}(\gamma_\agg) > 
\overline{p_c}(\gamma_\agg')$ for 
$\gamma_\agg < \gamma_\agg'$.
Here, $\overline{p_c}(\gamma_\agg)$ represents the statistical
average of $p_c(\gamma_\agg, i_s)$'s as a function of $\gamma_\agg$.
In fact, Figure \ref{fig:thandexvsag} shows that
$\overline{p_c}(\gamma_\agg) > \overline{p_c}(\gamma_\agg')$
for $\gamma_\agg < \gamma_\agg'$. 

When an individual $p_c(\gamma_\agg, i_s)$ becomes smaller than 
$\overline{p_c}(\gamma_\agg)$ for a large $\gamma_\agg$, the non-vanishing 
total conductivity $\sigma_\total(p,\gamma_\agg,i_s)$ 
is very small at $p \in (p_c(\gamma_\agg,i_s), \overline{p_c}(\gamma_\agg))$
and increases weakly with respect to $p$ there.
It means that the exponent $\talpha(\gamma_\agg, i_s)$ 
becomes large in the conductivity curve (\ref{eq:sigmatotal}).
Figure \ref{fig:expvsthr}(b) illustrates the correlation between the exponent
$\talpha(\gamma_\agg, i_s)$ and the threshold $p_c(\gamma_\agg, i_s)$ 
in ACPMs, which means that the smaller
the thresholds $p_c(\gamma_\agg, i_s)$ are, the larger the exponents 
$\talpha(\gamma_\agg, i_s)$ are. 
These properties are the same as Figure \ref{fig:expvsthr}(a).

However the range of Figure \ref{fig:expvsthr}(b) quite differs from
Figure \ref{fig:expvsthr}(a). The 
%\textcolor{red}{\sout{fluctuations}}
%\textcolor{blue}{
variances
%}
in Figure \ref{fig:largeparticle} look symmetric with respect to their
averages whereas those in Figure \ref{fig:thandexvsag} are
asymmetry.
The larger $\gamma_\agg$ is, the smaller the average of thresholds 
$\overline{p_c}(\gamma_\agg)$ is and
the larger that of the exponents $\overline{\talpha}(\gamma_\agg)$ is. 
Thus the trend of Figure \ref{fig:thandexvsag} could
not be interpreted only by the size effect. The difference 
%\textcolor{red}{\sout{should}}
%\textcolor{blue}{
might
%}
 be regarded as an shape effect of the agglomerated clusters.
%}

In the previous work \cite{MSW}, we investigated the shape effect on
the CPMs for spheroid. 
The thinner the spheroid (of oblate case)
is, the smaller we have thresholds and
the larger we obtain exponents,
whereas the thicker the spheroid (of prolate case) is,
the smaller the threshold and the exponent are\cite{MSW,GSDT} .
As mentioned in Ref.~\cite{MSW}, the shape effect can be also
interpreted as the broad distribution continuum percolation
model (BDCPM). Since the spheroids with random orientation can be regarded
as a distribution (of probability) of the local conductivity,
CPMs for a shaped object, e.g., spheroid can be
interpreted as BDCPM from view point of the  probability theory.
It means that such properties of CPMs of the spheroid can be
applied to any shape problem in CPM including this case.
The thinner the shape of particles (or clusters) is, 
the smaller the threshold is and the larger the exponent is.

As showed in Subsection \ref{subsubsec:4.1.2},
it could be regarded that the larger  $\gamma_\agg$ is,
the thinner the shape of the agglomerated cluster becomes.
Hence the properties of the averaged values in
Figure \ref{fig:thandexvsag} 
%\textcolor{red}{\sout{can}}
%\textcolor{blue}{
could
%}
 be interpreted
as the shape effects.
In other words, the tendency of the threshold and
the exponent agrees with that of thin shape effects of the 
spheroids.
With the arguments in 
Subsections \ref{subsubsec:4.1.1} and \ref{subsubsec:4.1.2},
it means that the larger $\gamma_\agg$ is, 
the smaller 
%\textcolor{red}{\sout{
%we have the threshold and the larger we obtain the exponent.
%}}
%\textcolor{blue}{
the threshold is and the larger the exponent is.
%}
This property reproduces Figure \ref{fig:thandexvsag} and
Table \ref{tbl:thandexvsag}.

\subsubsection{The 
%\textcolor{red}{\sout{universality on}}
% \textcolor{blue}{
universal properties of
%}
conductivity in ACPM}
\label{sec:univACPM}

The 
%\textcolor{red}{\sout{fluctuations}}
%\textcolor{blue}{
variances
%}
of the threshold in CPM
are essentially the same as those in  ACPM  
if the agglomeration parameter $\gamma_\agg$ 
reads the size $\rho$ appropriately.
Though the dependence of the average of the exponents on
$\gamma_\agg$ differs from the $\rho$ dependence,
we consider that the difference might come from the fact that
the agglomerated cluster has a thin shape as another geometrical
effect.
%}

Let us 
%\textcolor{red}{\sout{consider}}
%\textcolor{blue}{
attempt an investigation of
%} 
the universal properties of ACPMs or the
behavior of ACPMs in the infinite region.

\begin{figure}[htbp]
 \begin{tabular}{cc}
  \begin{minipage}[t]{0.5\hsize}
   \begin{center}
    \includegraphics[width=\hsize, clip]{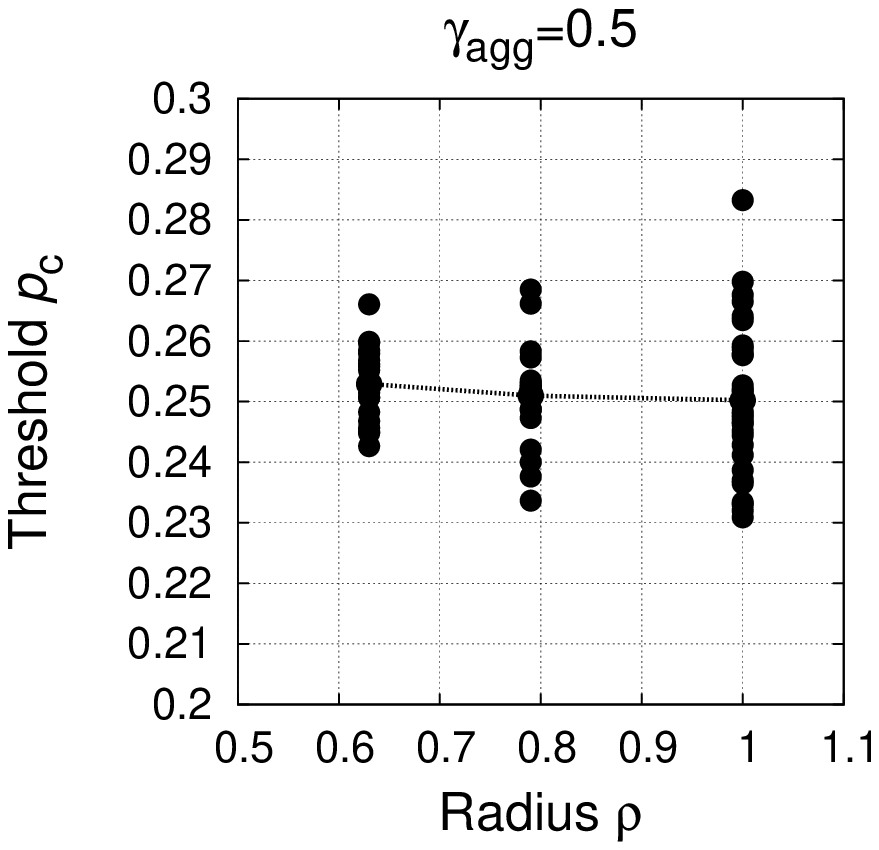}
    (a)
   \end{center}
  \end{minipage} 

  \begin{minipage}[t]{0.5\hsize}
   \begin{center}
    \includegraphics[width=\hsize, clip]{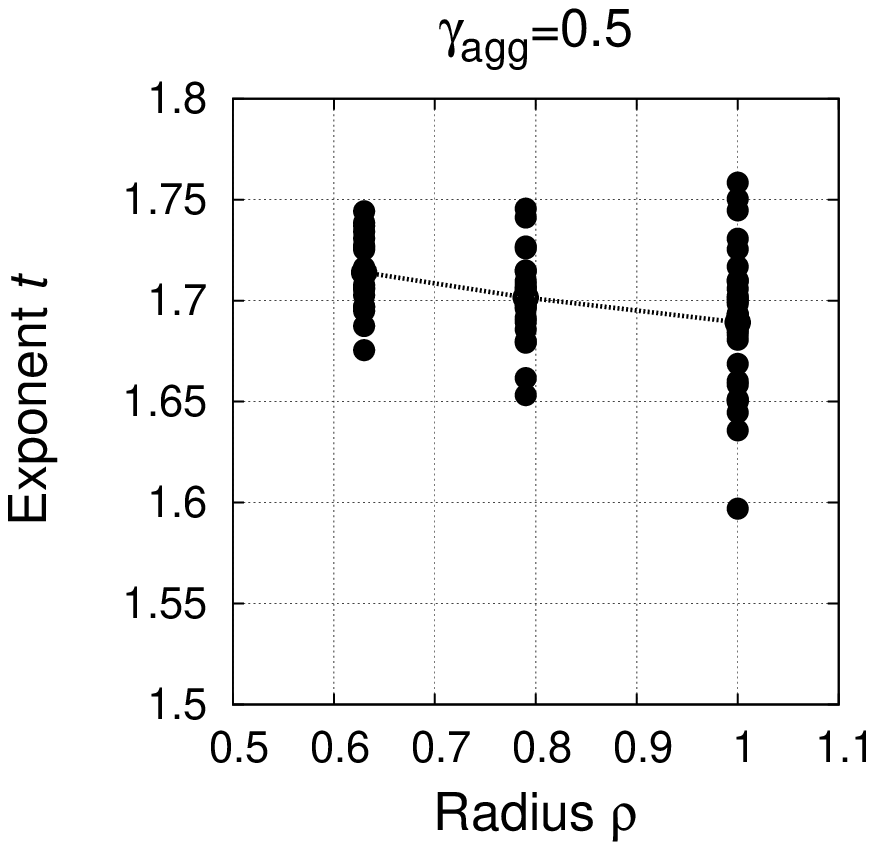}
    (b)
   \end{center}
  \end{minipage} 
 \end{tabular}
\caption{\label{fig:smallparticle}
The dependence of the thresholds and exponents on the
size of the system for ACPM with $\gamma_\agg=0.5$.
The filled circle corresponds to a computation for each seed
$i_s$ 
%\textcolor{blue}{
of the pseudo-randomness
%} 
and the dotted line is their average.
}
\end{figure}

Figure \ref{fig:smallparticle} exhibits the dependence of 
the threshold and the exponent on the size of the system 
for the $\gamma_\agg=0.5$ case.
In the computations of Figure \ref{fig:smallparticle}, we used larger
$N=N_x=N_y=N_z$ analyzed regions as $N=272$ and $N=344$.
The relative particle sizes (the radii) are $\rho=0.79$ and $\rho=0.63$.
Since these computations are harder than $N=216$, we
computed only twenty curves for these additional cases respectively
as shown in Figure \ref{fig:smallparticle} and Table \ref{tbl:smallparticle}.
 
\begin{table}[htbp]
\begin{center}
\caption{The $\rho$ dependence
of the threshold and the exponent for the case $\gamma_\agg=0.5$.}
\label{tbl:smallparticle}
\begin{tabular}{|c|c|c|c|c|c|c|c|c|c|c|}
\hline
\multicolumn{1}{|c|}{ } & 
\multicolumn{3}{|c|}{Threshold}& 
\multicolumn{3}{|c|}{Exponent} \\
\hline
$\rho$         & Average & Maximum & Minimum & Average & Maximum & Minimum \\
\hline
$1.00^\dagger$ & 0.250   & 0.283   & 0.231   & 1.689   & 1.758   & 1.597   \\
$0.79^*$       & 0.251   & 0.269   & 0.234   & 1.702   & 1.746   & 1.653   \\
$0.63^*$       & 0.253   & 0.266   & 0.243   & 1.714   & 1.744   & 1.676   \\
\hline
\end{tabular}

$*$: twenty curves. $\dagger$ thirty curves.
\end{center}
\end{table}

Figure \ref{fig:smallparticle} and Table \ref{tbl:smallparticle}
show that the smaller the radius $\rho$ is, the smaller the 
%\textcolor{red}{\sout{fluctuations}}
%\textcolor{blue}{
variances
%} 
are.
This means that the 
%\textcolor{red}{\sout{fluctuations}}
%\textcolor{blue}{
variances
%}
come from the size effect.
They also show that the asymptotic values of the thresholds $p_c$
and the exponents $t$ might differ from those of $\gamma_\agg = 0.0$, and
they might not approach to those of $\gamma_\agg = 0.0$.
We conjecture that the agglomeration, 
at least in the case of our algorithm of agglomeration,
%\textcolor{red}{\sout{has}} \textcolor{blue}{
would have
%}
 effects on 
%\textcolor{red}{\sout{universality class.}}
% \textcolor{blue}{
the universal properties.
%}
%\textcolor{blue}{
If there is the effect,
the difference is expected to come from the shape effect of the agglomerated
clusters as argued above.
%}

%\begin{acknowledgments}
%We wish to acknowledge the support of the author community in using
%REV\TeX{}, offering suggestions and encouragement, testing new versions,
%\end{acknowledgments}

\section{Summary}

By employing the simple algorithm to simulate an agglomerated configuration
of particles, which is controlled by the agglomeration parameter
$\gamma_\agg \in [0,1]$,
we numerically show how the thresholds $p_c$ and the exponent $\talpha$
of the conductivity curve 
(\ref{eq:sigmatotal}) depend upon the agglomerations
as shown in Figure \ref{fig:thandexvsag}.
%\textcolor{blue}{
Since our algorithm is given as a sequence of conditional probabilistic
events (\ref{eq:FiltR}), we can statistically investigate
the conductivity curves as functions over the sequences.
%}
The larger 
the agglomeration parameter $\gamma_\agg$ is,
the larger the 
%\textcolor{red}{\sout{fluctuations are}.}
%\textcolor{blue}{
variance is.
%}
Further the larger $\gamma_\agg$ is,
the smaller the average of the thresholds is and
the larger the average of the exponents is.

From Section 4, we conclude that the origin of these effects could
be interpreted as the size effect mainly,
because the agglomeration makes the percolation clusters larger
than those of non-agglomerated case.
Since for the finite region, 
the enlargement of the clusters
makes the 
%\textcolor{red}{\sout{fluctuations}}
%\textcolor{blue}{
variances
%}
of the conductivity large,
the size effect is crucial if the size of system is not
large enough.

Subsidiarily 
%\textcolor{blue}{
it is expected that
%}
the shape of the agglomerated
clusters also affect the conductivity.
It means that the shape of the agglomerated clusters
%\textcolor{red}{\sout{has}} \textcolor{blue}{
might have
%}
 effects on 
%\textcolor{red}{\sout{the universality class of }}
% \textcolor{blue}{
the universal properties of
%}
the threshold and the exponents 
as 
%\textcolor{red}{\sout{shown in}} 
%\textcolor{blue}{
we attempted an investigation in
%}
Subsection \ref{sec:univACPM}.

Due to the development of the technology,
devices with small sizes and the small conductive nanoparticles are concerned.
Small particles basically agglomerate much due to their surface energy.
Then in the conductivity in the composite materials of
the conductive nanoparticles,
the agglomeration causes the 
%\textcolor{red}{\sout{fluctuations}}
%\textcolor{blue}{
variances
%}
of the conductivity and the difference of individual devices.
Hence we believe that our results shed a new light on the applications of
the percolation theory to such a real material system.

%\textcolor{blue}{
\section*{acknowledgment}
The authors are grateful to anonymous referees for their
valuable suggestions and helpful comments.
%}



%% References


%\bibliographystyle{elsarticle-num}



\end{document}